\documentclass{aa} 
\usepackage{float} 
\usepackage{graphicx}  
\usepackage{graphics}
\usepackage{txfonts}  
\usepackage{natbib}
\usepackage{float}
\bibpunct{(}{)}{;}{a}{}{,}
\usepackage{color}
\usepackage[normalem]{ulem}

\begin{document}  

% \bibpunct{(}{)}{;}{a}{}{,} % to follow the A&A style 
 
\title {Chemical mixing in low  mass stars:\\ \Large{ I. Rotation against atomic diffusion including radiative accelerations}}

  \author{M. Deal\inst{1,2}, 
        M-J. Goupil\inst{2},
        J. P. Marques\inst{3},
        D. Reese\inst{2}
          \and
        Y. Lebreton \inst{4,2}}
  
\institute{Instituto de Astrofísica e Ciências do Espaço, Universidade do Porto
CAUP, Rua das Estrelas, PT4150-762 Porto, Portugal     
           \and
           LESIA,  Observatoire  de  Paris,  PSL  Research  University,  CNRS, Sorbonne Universit\'es, UPMC Univ. Paris 06, Univ. Paris Diderot,Sorbonne Paris Cit\'e, 92195 Meudon Cedex, France
           \and
           Institut d’Astrophysique Spatiale, UMR8617, CNRS, Universit\'e Paris XI, B\^atiment 121, 91405 Orsay Cedex, France
		   \and
           Univ Rennes, CNRS, IPR (Institut de Physique de Rennes) - UMR 6251, F-35000 Rennes, France\\
            \email{morgan.deal@astro.up.pt} 
           }
           
\date{\today}

\abstract
% Context
{When modelling stars with masses larger than 1.2~M$_{\odot}$ with no observed chemical peculiarity, atomic diffusion is often neglected because, on its own, it causes unrealistic surface abundances compared with those observed. The reality is that atomic diffusion is in competition with other transport processes. Rotation is one of the processes able to prevent excessively strong surface abundance variations.}
% Aims
{The purpose of this study is to quantify the opposite or conjugated effects of atomic diffusion (including radiative accelerations) and rotationally induced mixing in stellar models of low mass stars, and to assess whether rotational mixing is able to prevent the strong abundance variations induced by atomic diffusion in F-type stars. Our second goal is to estimate the impact of neglecting both rotational mixing and atomic diffusion in stellar parameter inferences for stars with masses larger than 1.3~M$_{\odot}$.}
% Method
{Using the AIMS stellar parameter inference code, we infer the masses and ages of a set of representative artificial stars for which models were computed with the CESTAM evolution code, taking into account rotationally induced mixing and atomic diffusion, including radiative accelerations. The 'observed' constraints are asteroseismic and classical properties. The grid of stellar models used for the optimisation search include neither atomic diffusion nor rotationally induced mixing. The differences between real and retrieved parameters then provides an estimate of the errors made when neglecting transport processes in stellar parameter inference.}
% Results
{We show that for masses lower than 1.3~M$_{\odot}$, rotation dominates the transport of chemical elements, and strongly reduces the effect of atomic diffusion, with net surface abundance modifications similar to solar ones.
At larger mass, atomic diffusion and rotation are competing equally. Above 1.44~M$_{\odot}$, atomic diffusion dominates in stellar models with initial rotation smaller than $80~\mathrm{km.s^{-1}}$ producing a chemical peculiarity which is not observed in \textit{Kepler}-legacy stars. This indicates that a transport process of chemical elements is missing, probably linked to the missing transport process of angular momentum needed to explain rotation profiles in solar-like stars. Importantly, neglecting rotation and atomic diffusion (including radiative accelerations) in the models, when inferring the parameters of F-type stars, may lead to errors of $\approx5\%$, $\approx2.5\%$ and $\approx25\%$ respectively for stellar masses, radii and ages.}
% Conclusions
{Atomic diffusion (including radiative accelerations) and rotational mixing should be taken into account in stellar models in order to determine accurate stellar parameters. When atomic diffusion and shellular rotation are both included, they enable stellar evolution codes to reproduce the observed metal and helium surface abundances for stars with masses up to 1.4~M$_{\odot}$ at solar metallicity. However, for these stars, if rotation is actually uniform, as observations seem to indicate then an additional chemical mixing process is needed together with a revised formulation of rotational mixing. For larger masses, an additional mixing process is needed in any case.}

  
\titlerunning{Rotation against atomic diffusion}
  
\authorrunning{Deal et al.}  

\maketitle 

%_____________________________________________________________________
%%%%%%%%%%%%%%%%%%%%%%%%%%%%%%%%%%%%%%%%%%%%%%%%%%%%%%%%%%%%%%%%%%%
\section{Introduction}

The determination of stellar classical parameters (such as age, mass and radius) is possible through the computation of stellar models. These models depend on micro and macro-physics inputs ingredients and can be constrained thanks to spectroscopic, astrometric, photometric, interferometric, and asteroseismic observations. Among these, asteroseismology is the only one that can give information on the internal structure of stars, the others giving access to surface layers only.

In order to reproduce all available observations, the physical processes at work in the stellar plasma have to be identified and, if recognized as efficient, have to be taken into account in stellar models. This can imply a heavy computational cost, but is a necessary step towards the determination of  stellar parameters as accurate as possible . In this respect, chemical element transport in stellar interiors is a key ingredient. While macroscopic transport by convection has been recognized as a major process a long time ago, microscopic transport due to gravitational settling has been considered to be a crucial transport process only after helioseismology placed constraints on the solar sound speed profile \citep{christensen93}.

Transport processes of chemical elements are of two types: microscopic (hereafter atomic diffusion) and macroscopic processes. The origin of atomic diffusion comes from the first principles of physics and is a direct consequence of the fact that stars are self-gravitating spheres of plasma composed of different gases (chemical element mixture). The equilibrium of stars leads to internal gradients of pressure, temperature, density and a radiative transfer producing a selective motion of chemical elements. Atomic diffusion is mainly led by the competition between gravity, which makes the elements move toward the center of the star, and radiative acceleration, which is due to the transfer of momentum between photons and ions and makes the elements move toward the surface. The combination of both processes leads either to a local depletion of elements or to an accumulation, the net effect depending on the chemical element under consideration. 

Macroscopic transport processes are plasma instabilities which induce a transport of chemical elements on a large scale. These processes are most of the time treated as diffusive and are in direct competition with atomic diffusion. This competition occurs for all types of stars and at each evolutionary stage. The result of this competition depends on the relative time scales of all possible processes. The time scale of atomic diffusion for stars with masses close to the solar one is long (5-10 Gyrs) and decreases for more massive stars (up to a dozen Myrs for A and B type stars). Depending on the competing transport processes, the variations of the surface abundances are different. If the atomic diffusion time scale is longer or is shorter than that of a competing transport process, then the surface abundances remain close to the birth composition of the star. This is the case of the Sun where a 10\% depletion of helium and metals is observed \citep{bahcall92,bahcall95,christensen96,richard96,ciacio97,gabriel97,morel97,brun98,elliott98,turcotte98}. In other cases when the atomic diffusion time scale is short, surface abundances can be very different from the initial ones, as in chemically peculiar stars \citep[e.g][]{praderie67,michaud70,watson70,watson71,richer00,richard01,michaud11,deal16}. In standard models the only macroscopic transport process included is convection. In convective zones, the mixing of chemical elements is very efficient and we can consider that atomic diffusion is negligible. But convection alone is not enough to explain the surface abundances of observed stars. It is then mandatory to include non standard processes and one of the most studied ones is the mixing of chemical elements induced by the rotation of the star.

\cite{michaud82} first quantified the competition between atomic diffusion and meridional circulation in the envelopes of AmFm stars. Two-dimensional simulations of this competition were carried-out by \cite{charbonneau88, charbonneau91}. These authors assumed a quasi-solid body rotation and an advective transport of chemical elements by meridional circulation \citep{tassoul82}. They concluded that AmFm stars can exist if rotation is slower than 100~km/s\, meaning that in these stars atomic diffusion is the dominant transport process. It is worth pointing out that AmFm stars are at the border of the solar-like oscillating star (K7-F5) domain that we consider in this work.

Rotation is often invoked as the process that prevents atomic diffusion to have a significant impact in stars, as it is able to efficiently transport chemical elements. A first aim of the present work is to check whether this assertion is valid  for masses in the range [1.3;1.44]~M$_\odot$ on the main sequence. For that purpose, we  built stellar models taking into account consistently both processes as they are currently modeled. We examined how different levels of rotation can counteract atomic diffusion processes. We first compared models of virtual stars and then examined how, in real stars, the different effects can be evidenced on the basis of observational constraints.

Asteroseismology allows us to probe the structure of stars. In particular, asteroseismic studies demonstrated that atomic diffusion has an impact on the inner structure of stars \citep[see e.g.][]{deal17,deal18}.  \cite{eggenberger10} investigated the combined impact of rotation and atomic diffusion for 1~M$_\odot$ models. The authors showed for instance that rotation by counteracting atomic diffusion keeps the surface helium from decreasing significantly compared with the case of non-rotating models. Some effects on specific seismic diagnostics nevertheless exist due to the structural differences between models without rotation and those with rotation and atomic diffusion. As we will see hereafter, this remains true as long as the impact of rotation dominates over the impact of atomic diffusion that-is for stars with small masses. It is then necessary to quantify the impact of atomic diffusion on the stellar parameter determination for stars with increasing masses.

The combined effect of atomic diffusion including radiative accelerations and rotational transport on star surface abundances and internal structures has never been studied in the light of the very precise asteroseismic data available nowadays from \textit{Kepler} and, the new generation of stellar models including atomic diffusion and the effect of rotation. We use the CESTAM evolution code \citep{morel08,marques13,deal18} which is one of the few evolution codes (together with the TGEC code \citealt{hui-bon-hoa08,theado12} and MESA \citealt{paxton18}) able to compute such models self-consistently. The errors we make on the stellar parameter inference when   neglecting rotation and/or atomic diffusion are then estimated using the AIMS (Asteroseismic Inference on a Massive Scale) stellar model optimization pipeline \citep{rendle19}.

A second goal of the present paper, is to assess the net result of atomic diffusion (including radiative accelerations) and rotationally induced mixing on the surface abundance and stellar parameter determination of solar-like oscillating stars with masses below 1.5~M$_{\odot}$. For that purpose, stellar models were built and are described in Section 2. The impact of atomic diffusion and rotation on the transport of chemical elements is presented in Section 3. The method we used to infer stellar parameters and the associated grid of models are described in Section 4. Then the results of optimizations are presented in Section 5. Finally we discuss about the results and conclude in Section 6.

\section{Stellar models}

We computed stellar models with the CESTAM evolution code \citep{morel08,marques13}. We focus on the interval of mass where the effect of radiative accelerations is important, i.e. between $1.3$ and $1.44$~M$_{\odot}$ \citep{deal18}, and, where the efficiency of the transport of chemical elements by rotation is decreasing \citep[e.g.][]{talon08}. In this exploratory work, due to the important amount of computational time needed, we only consider models at solar metallicity. 

\subsection{Input physics, including atomic diffusion}

In this study, the important input physics are atomic diffusion, as well as opacities which are affected by the modification of chemical mixture induced by diffusion.

Atomic diffusion was computed following the \cite{michaud93} formalism and the Single-Valued Approximation \citep{leblanc04} for radiative acceleration. Details on the atomic diffusion computation in CESTAM are given in \cite{deal18}. The elements and isotopes that are followed are $^1$H, $^3$He, $^4$He, $^{12}$C, $^{13}$C, $^{14}$N, $^{15}$N, $^{15}$O, $^{16}$O, $^{17}$O, $^{22}$Ne, $^{23}$Na, $^{24}$Mg, $^{27}$Al, $^{28}$Si, $^{31}$P (without radiative accelerations), $^{32}$S, $^{40}$Ca, and $^{56}$Fe.

Rigourously speaking, the Rosseland mean opacities should be computed for the exact -varying- chemical mixture at each mesh point in the star and each evolution time step. One then could use the OPCD3 package provided by the Opacity Project\footnote{http://cdsweb.u-strasbg.fr/topbase/testop/TheOP.html}. However, such opacity calculations are excessively time consuming. Therefore, we restricted the full computation of opacity with OPCD3 to the interval $T\in [10^4, 10^{6.23}]$~K. In this temperature range, we found that the contribution to opacity of the elements that vary the most due to atomic diffusion (mainly iron here) is important. This requires a detailed opacity calculation for the exact associated chemical mixture. On the other hand, in the regions where T$>10^{6.23}$~K and T$<10^4$~K, the initial solar mixture is only slightly affected by atomic diffusion and the impact on opacity can safely be neglected. There, we used the opacity tables of the Opacity Project \citep[OP,][]{seaton05} and the Wichita opacity data \citep{ferguson05} respectively. We checked that the opacity smoothly varies at the boundaries of the different tables. Both sets of tables correspond to a fixed solar mixture, the \citet{asplund09} mixture with meteoritic abundances for refractory elements as recommended by \cite{serenelli10}. 

Convection is treated following \cite{canuto96}'s prescription. The mixing-length parameter is calibrated so that the solar model reaches the solar radius and luminosity at solar age which yields $\alpha_\mathrm{CGM}=0.68$ \citep[see e.g.][]{lebreton14} for models with atomic diffusion and without rotation. We chose to keep this value for all the models of the study in order to assess the impact of the sole  physical process under  study. We discuss this point in Section \ref{alpha}.

Other input physics are the same as used by  \citet[][]{lebreton14}. Briefly, we used the OPAL2005 equation of state \citep{rogers02} and nuclear reactions rates from the NACRE compilation \citep{angulo99} except for the $^{14}\mathrm{N}(p,\gamma)^{15}\mathrm{O}$ reaction, for which we use the LUNA rate \citep{imbriani04}. Atmospheres are computed in the grey approximation and integrated up to an optical depth of $\tau = 10^{-4}$ and we do not consider mass loss.

\subsection{Rotational mixing}\label{theo_rota}

Rotationally induced transport as implemented in CESTAM is detailed in \cite{marques13}. The transport of angular momentum obeys the advection-diffusion equation:
\begin{equation}
 \frac{\mathrm{d}}{\mathrm{d}t}\left(r^2\Omega\right)_{M_r}=\frac{1}{5 \rho r^2}\frac{\partial}{\partial r}\left(\rho r^4 \Omega U_2\right) + \frac{1}{ \rho r^2}\frac{\partial}{\partial r} \left(\rho \nu_v r^4 \frac{\partial \Omega}{\partial r}\right)
\end{equation}
\noindent  where $\rho$ is the density, $r$ is the local radius, $\Omega$ the mean angular velocity at radius $r$, $U_2$ the vertical component of meridional circulation velocity at radius $r$ and $\nu_v$ the vertical component of the turbulent viscosity. 

When angular momentum loss is taken into account at the surface, we used the prescription of \cite{matt15} for magnetized wind braking with a solar calibrated constant of $1.25\times10^{31}$~erg (except in the uniform rotation case, see Section \ref{solid}). Models are computed from the ZAMS with an initial surface rotation of either 30 km/s or 80 km/s to cover the range of rotation speed observed for \textit{Kepler} stars showing detectable solar-like oscillation frequencies.

The vertical component of the turbulent diffusivity is given by \citep{chaboyer92,talon97}: 
\begin{equation}\label{Dv}
D_v=\frac{Ri_c(K+D_h)r^2}{N_T^2+N^2_\mu (1+K/D_h)}\left(\frac{\partial\Omega}{\partial r}\right)^2
\end{equation}

\noindent where $Ri_c=1/6$ is the critical Richardson number, $K$ the thermal diffusivity, $N_T$ and $N_\mu$ the chemical and thermal parts of the Brunt-Väisälä frequency, and $D_{h}$ the horizontal component of turbulent diffusivity. For $D_h$, we used the prescription of \citet{mathis04}:
\begin{equation}
D_{h}=\sqrt{\frac{\beta}{10}r^3 \Omega |2V_2-\alpha U_2|},
\end{equation}
where $V_2$ and $U_2$ are the horizontal and vertical components of the velocity of the meridional circulation, $\alpha=\frac{1}{2}\frac{\partial \ln r^2 \Omega}{\partial \ln r}$ and $\beta=1.5\times10^{-5}$ \citep{richard99}.
We assumed that the vertical turbulent diffusivity $D_V$ is equal to the vertical turbulent viscosity $\nu_V$.

The combination of the transport of chemical elements by meridional circulation and horizontal turbulence results in a diffusive process with a coefficient given by \citet{chaboyer92}:
\begin{equation}
{D_\mathrm{eff}}=\frac{(r U_2)^2}{30 D_h}.
\end{equation}
The total diffusion coefficient of chemical elements associated with rotation is then given by
\begin{equation}\label{dturb}
{D_\mathrm{turb,rota}}=D_v+D_{\rm eff}.
\end{equation}
 
This formalism for transport of angular momentum was used to compute the evolution of the rotation profile of stellar models for solar-like oscillating stars. The seismic diagnostics associated with the resulting rotation profiles were then compared with the observational ones \citep[e.g.][]{eggenberger12,marques13}. It was shown to be too inefficient to account for the internal rotation inferred from observed stars \citep[e.g][]{deheuvels12,aerts18,ouazzani19,buldgen19} despite various efforts for improvement \cite[see e.g.][]{mathis18}. The purpose of the present paper is not to reproduce real stars, but rather to understand the interaction between microscopic and macroscopic aspects of the transport. It is a step forward to a more detailed and accurate theory for the transport of chemical elements in stars. Nevertheless we also investigate the case when a uniform rotation is imposed during the evolution of the stellar models.

\subsection{Coupling between atomic diffusion and rotational mixing}\label{coupling}

The interaction between rotation and atomic diffusion occurs at different levels. One of them is the impact of the $\mu$-gradient on the meridional circulation \citep[e.g.][]{charbonnel98} which will not be discussed in this study. We rather focus on the most obvious one that is the direct competition between atomic diffusion and turbulent diffusion (see Eq. \ref{dturb}). This coupling clearly appears in the diffusion equation:

\begin{equation}\label{equadiff}
\rho\frac{\partial X_i}{\partial t}=\frac{1}{r^2}\frac{\partial}{\partial r}\left[r^2\rho {D_\mathrm{turb,rota}}\frac{\partial X_i}{\partial r}\right]-\frac{1}{r^2}\frac{\partial}{\partial r}[r^2\rho v_{i}] + A_i m_p \left[\sum_j (r_{ji} -  r_{ij}) \right],
\end{equation}

\noindent where $X_i$ is the mass fraction of element $i$, $v_i$ its atomic diffusion velocity, $\rho$ the density in the considered layer, $D_\mathrm{turb,rota}$ the turbulent diffusion 
coefficient, $A_i$ its atomic mass, $m_p$ the mass of a proton, and $r_{ij}$ the 
reaction rate of the reaction that transforms element $i$ into $j$.

In the case of a trace element $i$, $v_i$ can be expressed  as: 

\begin{equation}\label{eqvdiff}
v_i=D_{ip}\left[-\frac{\partial \ln X_i}{\partial r}+\frac{A_i m_p}{k T}(g_{rad,i}-g)+\frac{({\bar{Z}_i}+1)m_p g}{2 k T}+\kappa_T\frac{\partial \ln T}{\partial r}\right],
\end{equation}

\noindent where $D_{ip}$ is the diffusion coefficient of element $i$ relative to protons. The variable $g_{rad,i}$ is the radiative acceleration on element $i$, $g$ the local gravity, $\bar{Z}_i$ the average charge (in proton charge units) of element $i$ (roughly equal to the charge of the `dominant ion'), $k$ the Boltzmann constant, $T$ the temperature, and $\kappa_T$ the thermal diffusivity.

Regarding Equation \ref{equadiff}, we can define three regimes:
\begin{itemize}
\item[1.] ${D_\mathrm{turb,rota}}\times\frac{\partial X_i}{\partial r} \gg v_i$, when the rotation is strong enough to significantly reduce the effect of atomic diffusion
\item[2.] ${D_\mathrm{turb,rota}}\times\frac{\partial X_i}{\partial r} \approx v_i$, when rotation and atomic diffusion are of the same order of magnitude
\item[3.] ${D_\mathrm{turb,rota}}\times\frac{\partial X_i}{\partial r} \ll v_i$, when the rotation is not able to significantly reduce the effect of atomic diffusion.
\end{itemize}

As will be shown in next section, the three regimes are encountered in solar-like oscillating main-sequence stars. The value of $D_\mathrm{turb,rota}$ -- and therefore the efficiency of the rotationally induced mixing -- depends mainly on the value of $D_v$ as $D_h \gg (rU_2)^2/30$ in all the  models considered in this study (see Eq. \ref{dturb}) . 
On the other hand, $D_v$ mainly depends on the angular momentum gradient (Eq. \ref{Dv}). Hence rotation has a strong impact on the transport of chemical elements when the rotation velocity is high or when the gradient of angular momentum is large such as at the bottom of the surface convective zone. In the other regimes, the effect of rotation is negligible or of the same order of magnitude as atomic diffusion.

The presence of an angular momentum gradient at the bottom of a stellar surface convective zone and the associated mixing of chemical elements then depend  on how well the meridional circulation is driven. 
In the case when the total angular momentum is conserved, the efficiency of meridional circulation and the associated transport of chemical elements decrease rapidly with time. One of the rare (or few) processes able to maintain  meridional circulation efficiently on the main-sequence is the extraction of angular momentum at the surface during the evolution. This extraction produces a gradient of angular momentum which drives the meridional circulation (increasing the $U_2$ term) and then increases the vertical transport of angular momentum. This increase of the angular momentum transport leads to an important mixing of chemical elements able to reduce significantly atomic diffusion.

The most studied extraction of angular momentum at the surface of stars is the extraction through magnetized winds. Several empirical and theoretical prescriptions exist to model this extraction \citep[e.g][]{skumanich72,kawaler88,matt12,matt15}. This extraction is efficient for stars able to maintain a magnetic field generated by a dynamo. The star then needs a surface convective zone large enough to generate these fields. This is why the extraction of angular momentum due to magnetized winds is very efficient for stars with masses lower than 1.2~M$_{\odot}$ at solar metallicity. The efficiency is smaller when the mass of the star is larger, i.e. the size of the surface convective zone is smaller. This is what explains the hot part of the lithium dip \citep[e.g.][]{talon08}.

On the other hand, in regions of the star where radiative accelerations are efficient, atomic diffusion produces accumulations of elements. Accumulation of a chemical element occurs in layers where this element significantly contributes to the opacity. It results in a local increase of the opacity. In solar-like oscillating main-sequence stars the accumulation occurs at the surface which induces an increase of the size of the surface convection zone \citep{turcotte98b,deal17,deal18}. This in turn increases the efficiency of the extraction through magnetized wind and the mixing of chemical elements. But at the same time, the extraction of angular momentum makes the star rotate slower, decreasing the efficiency of mixing of chemical elements. The net result can therefore significantly differs from one star to another depending on its mass and evolutionary state.

In the next section we illustrate the aforementioned rotation-atomic diffusion interaction process with stellar evolution models computed with the current physical description of both rotation and atomic diffusion in a consistent way.

\section{Net mixing induced by rotation and atomic diffusion}\label{rota}

\begin{figure}[!h]
\center
\includegraphics[scale=0.58]{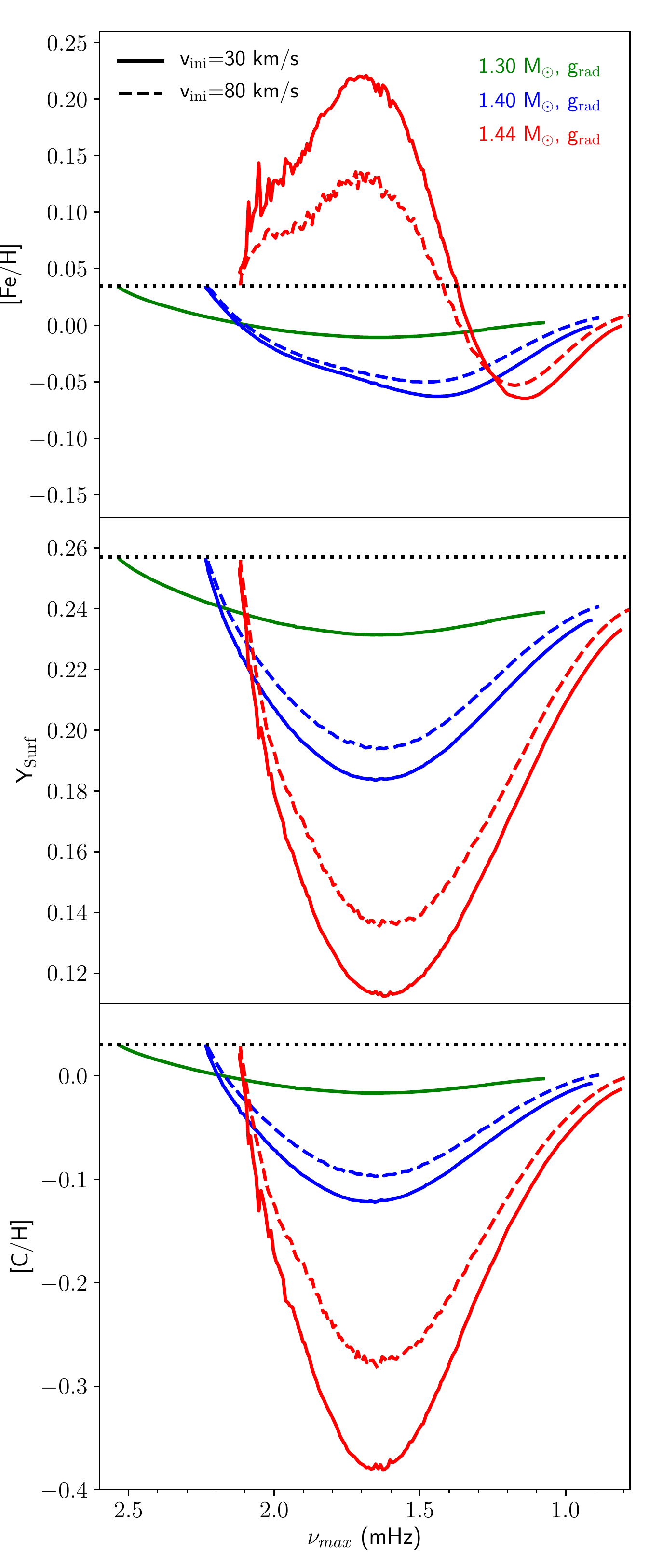}
\caption{Evolution of the surface [Fe/H] in dex (top panel), helium $Y_\mathrm{Surf}$ in mass fraction (middle panel) and the surface [C/H] in dex (lower panel) as a function of $\nu_\mathrm{max}$ for different masses and initial rotation speeds for models including atomic diffusion with radiative accelerations. Horizontal dotted lines represent the initial abundance. The models are A1, B1, B2, C1 and C2 from Table \ref{table:1}.}
\label{fig:1}
\end{figure}

\begin{figure}[!h]
\center
\includegraphics[scale=0.61]{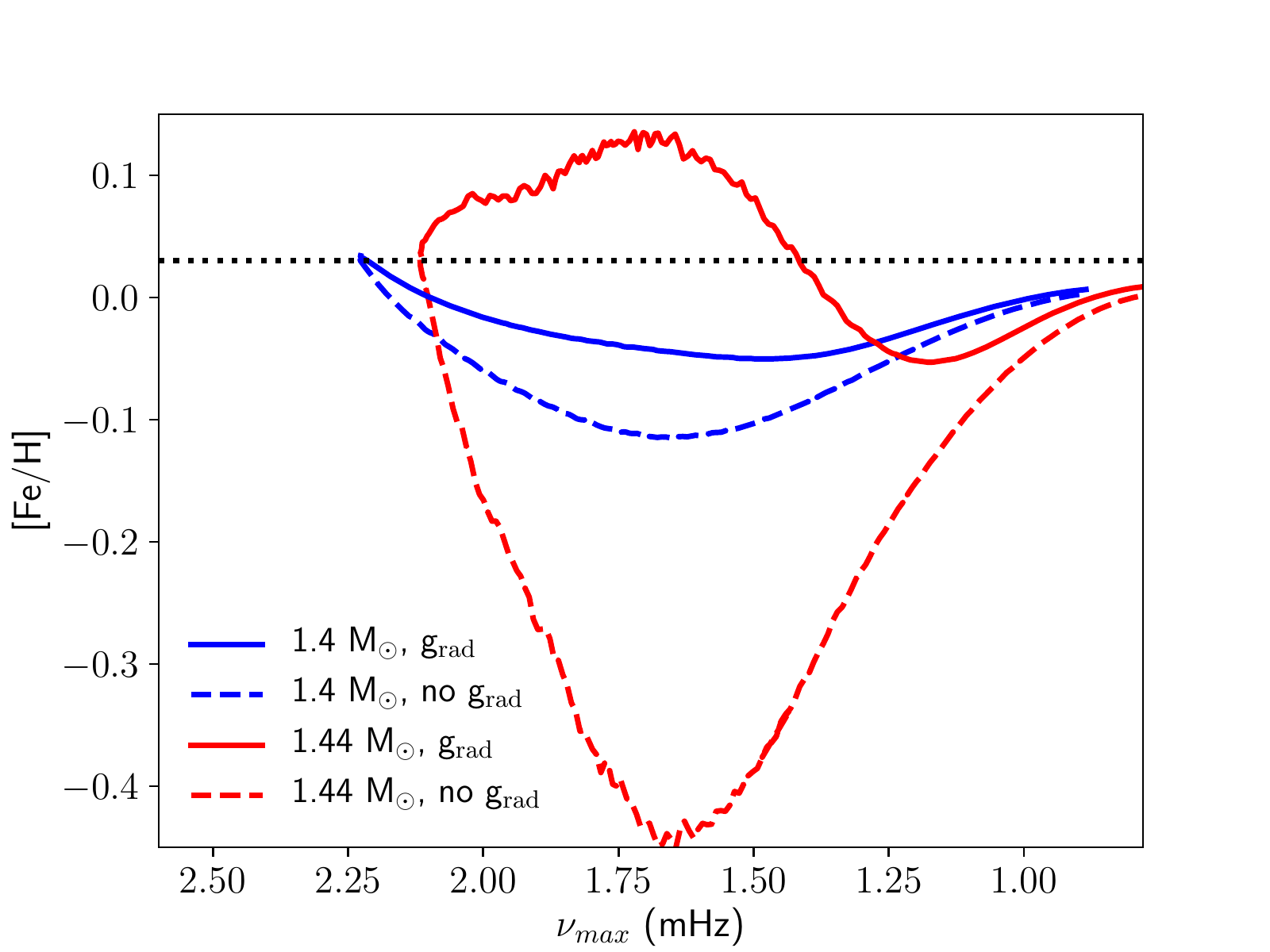}
\caption{Evolution of the surface [Fe/H] in dex with $\nu_\mathrm{max}$ for 1.4~M$_\odot$ models (B2 and B4) and 1.44~M$_\odot$ models (C2 and C4) with an initial rotation speed of 80~km/s. Models including atomic diffusion with (respectively without) radiative accelerations are shown with solid (respectively dashed) lines.}
\label{fig:1bis}
\end{figure}

\subsection{Surface abundances}
We showed in \cite{deal18} that the abundance variations produced by atomic diffusion alone can be very large and may not reproduce observations. This is because atomic diffusion does not occur alone in stars. There are other processes that are in competition with atomic diffusion and reduce its effects. Rotational mixing is one of them. Here we therefore aim at determining whether the surface abundances of solar-like oscillating main-sequence stars can be reproduced in models including simultaneously atomic diffusion and rotationally induced mixing.
To answer this question we computed stellar models with the characteristics listed in Table \ref{table:1}. We considered three values of the mass: $1.3, 1.4, 1.44\ M_\odot$ (models A,B,C respectively), which are found to correspond to the three regimes mentioned in section \ref{coupling}. The reference models (A1, B1, B2, C1 and C2) include rotationally induced mixing with two initial values of the rotation velocity, atomic diffusion with radiative acceleration, and extraction of angular momentum at the surface. We point out that the mixing induced by rotation is large enough to prevent numerical instabilities produced by radiative accelerations. We therefore do not have to impose an ad-hoc mixing at the base of the convective zone as is otherwise usually done \citep{theado09,deal16,deal18}.

\subsubsection{Impact of the initial rotational velocity}

Measured surface rotation rates of solar-like stars do not exceed 35~km/s \citep[e.g][]{benomar15}. This is why we concentrate on surface initial rotation velocities between 30 and 80 km/s in order to characterize the impact of atomic diffusion and rotation on stars that were observed (\textit{Kepler}, TESS) or will be observed (PLATO) for asteroseismology.

\begin{table*}
\centering
\caption{Inputs of the stellar models. All cases are not presented for all masses (for instance A2 and A4).} 
\label{table:1}
\begin{tabular}{l|cc|cccc|c|cccc|c}
\noalign{\smallskip}\hline\hline\noalign{\smallskip}
Model & A1 & A3 & B1 & B2 & B3 & B4 & B5 & C1 & C2 & C3 & C4 & C5\\
\noalign{\smallskip}\hline\noalign{\smallskip}
Mass (M$_{\odot}$) & 1.30 & 1.30 & 1.40 & 1.40 & 1.40 & 1.40 & 1.40 & 1.44 & 1.44  & 1.44 & 1.44 & 1.44\\ 
AM Loss & M15\tablefoottext{b} & M15 & M15 & M15 & M15 & M15 &  None & M15 & M15 & M15 & M15 & M15 \\
v$_{ini}$ (km/s) & 30 & 30 & 30 & 80 & 30 & 80 & 30 & 30 & 80 & 30 & 80 & 80\\
$g_\mathrm{rad}$. & yes & no & yes & yes & no & no & yes & yes & yes & no & no & yes \\
\noalign{\smallskip}\hline\noalign{\smallskip}
Opacities & \multicolumn{2}{c|}{OPCD+OP}& \multicolumn{5}{c|}{OPCD+OP}  &\multicolumn{4}{c|}{OPCD+OP}& OP\\
\noalign{\smallskip}\hline\noalign{\smallskip}
Rotation &  \multicolumn{12}{c}{TZ97\tablefoottext{a}}\\
X$_{ini}$ & \multicolumn{12}{c}{0.7280}  \\ 
Y$_{ini}$ & \multicolumn{12}{c}{0.2578} \\
Z/X$_{ini}$ & \multicolumn{12}{c}{0.0195} \\
$\alpha_\mathrm{CGM}$ & \multicolumn{12}{c}{0.68}  \\ 
Mixture & \multicolumn{12}{c}{AGSS09}  \\ 
EoS & \multicolumn{12}{c}{OPAL2005}  \\
Nuc. react. & \multicolumn{12}{c}{NACRE+LUNA} \\ 
Core ov. & \multicolumn{12}{c}{0.15} \\    
\noalign{\smallskip}\hline\noalign{\smallskip}
\end{tabular}
\tablefoot{\tablefoottext{a}{\cite{talon97}},\tablefoottext{b}{\cite{matt15}} }
\end{table*} 

Figure \ref{fig:1} shows the temporal evolution of the surface abundance of some elements (iron, helium and carbon). As an observed seismic proxy for the age, we use $\nu_\mathrm{max}$, the frequency at maximum power in the oscillation spectrum of solar-like pulsators. $\nu_\mathrm{max}$ indeed decreases as stars evolve on the main sequence \citep{brown91, kjeldsen95,kallinger10}. Regarding the models, we can identify three regimes (as presented in Section \ref{coupling}):

\begin{itemize}
\item For masses lower than 1.4~M$_\odot$ (illustrated by model A1 in Fig. \ref{fig:1}), we find that the effect of atomic diffusion is small compared with that of rotation. In this mass range, the extraction of angular momentum by the magnetized wind is efficient because the surface convective zone is still thick enough to maintain a dynamo. Meridional circulation is then still efficiently driven, and is able to maintain differential rotation. This induces a relatively strong mixing which prevents unrealistic chemical element depletions due to atomic diffusion. Accordingly, the surface abundance depletions are small and comparable to the ones observed in the Sun \citep[e.g.][]{bahcall92}. The helium and metal surface abundances resulting from the competition between atomic diffusion and rotation are compatible with observations \citep{ghazaryan18,verma19}, in particular the helium surface mass fraction $Y_s$ always remains higher than $0.18$.

We point out that even in this case the surface abundances vary in time. The surface variations are similar in the two models of 1.3~M$_\odot$ with initial rotation velocities of 30 km/s and 80 km/s (the latter, model A2, is not presented in Fig. \ref{fig:1}).

\item For stars with masses around 1.4~M$_\odot$ (see models B1 and B2 in  Fig. \ref{fig:1}), we find that atomic diffusion and rotation are equally important. Rotation prevents atomic diffusion from inducing large depletions/accumulations of metals and helium at the surface. Atomic diffusion still induces an iron depletion of 0.1~dex, which is of the  order of magnitude of the observed surface abundance uncertainties. Helium reaches a depletion level that is still consistent with observations \citep{ghazaryan18,verma19}. The initial rotation speed starts to have an effect on the surface abundances variations, but this effect remains too small to be observed.

\item For stars with masses larger than 1.4~M$_\odot$ (C1 and C2 models in Fig. \ref{fig:1}), we find that atomic diffusion dominates over rotation in the studied range of initial rotation speeds. Iron accumulation occurs at the surface while helium and carbon are strongly depleted. While the iron surface abundance is still compatible with observations, helium is too depleted \citep{ghazaryan18,verma19}. The effect of the initial rotation speed is now well visible in Fig. \ref{fig:1}, with a 0.1~dex decrease in the iron enhancement when rotation increases from 30 to 80 $\mathrm{km.s^{-1}}$. In this case the surface convective zone is thin ($\log(\Delta M/M^\ast)\approx-6)$), and an efficient dynamo cannot be maintained. This results in a less efficient extraction of angular momentum by magnetized wind. The meridional circulation is then not maintained and the mixing of chemical elements is weaker. On the other hand, atomic diffusion is very efficient close to the surface, so the thinner the surface convective zone is, the more efficient diffusion is. The combination of both effects leads to large surface abundance variations.

\end{itemize}

\subsubsection{Impact of radiative accelerations}

To quantify the effects of radiative accelerations, we now consider models A3, B3, B4, C3 and C4 which are identical to models A1, B1, B2, C1, C2 respectively, except that they do not include radiative accelerations. We identify again three regimes:
\begin{itemize}

\item We find that the effect of radiative accelerations is negligible for the 1.3 M$_\odot$ models (models A1 and A3 are similar) because their convective envelopes are too deep.

\item For the 1.4~M$_\odot$ model, we do not find any accumulation of iron due to radiative accelerations (but a moderate depletion) contrarily to what was obtained in \cite{deal18} for a 1.4~M$_\odot$ model at solar metallicity, but with no rotationnaly induced mixing (see their Fig. 1). This shows that rotationally induced mixing is strong enough to prevent the accumulation due to radiative accelerations. Moreover, the comparison of models that include radiative accelerations (B2) with models that do not (B4) in Fig. \ref{fig:1bis} shows that radiative accelerations do not counteract rotational mixing but mitigate the gravitational settling. As part of atomic diffusion, they alone induce an iron surface abundance difference reaching from 0.05~dex for an initial rotation speed $v_{ini}=$80~km/s to 0.1~dex for $v_{ini}=$30~km/s (not shown on the Figure).

\item At 1.44~M$_\odot$, we find different types of behaviour depending on whether radiative accelerations are taken into account or not: with radiative accelerations iron accumulates at the surface (model C2, Fig. \ref{fig:1bis}) while without them iron is strongly depleted (model C4). The difference reaches $\approx0.6$~dex. With or without radiative accelerations, as the surface convective zone at 1.44~M$_\odot$ is thin, atomic diffusion is efficient while the extraction of angular momentum is not. With or without radiative accelerations, atomic diffusion leads to strong surface abundance variations, much larger than what is expected from observations.

\end{itemize}

These results show how important it is to take into account all atomic diffusion processes is this range of mass ([1.4;1.44]~M$_\odot$) for a more coherent and realistic physical description. The fact that its effects may be in contradiction with observations points toward some missing additional mechanism in the stellar modelling. Atomic diffusion (including radiative accelerations) is crucial when the impact of competing transport processes is weak. 

The predicted large element depletions (at least for the studied elements) are not observed. Indeed the surface abundances of stars with thin surface convective zones show weak surface abundance depletion/accumulation. This indicates that there is a need for additional transport processes or a revised formulation of rotationnally induced transport in order to reproduce the observations. This conclusion was already reached when studying the impact of atomic diffusion alone. The modelling of this type of star required the inclusion of an ad-hoc turbulence \citep[e.g.][]{richer00,richard01,michaud11,verma19b} in order to be consistent with observations. This conclusion still holds when including rotationally induced mixing as implemented in CESTAM for these masses.

\begin{figure}
\center
\includegraphics[scale=0.5]{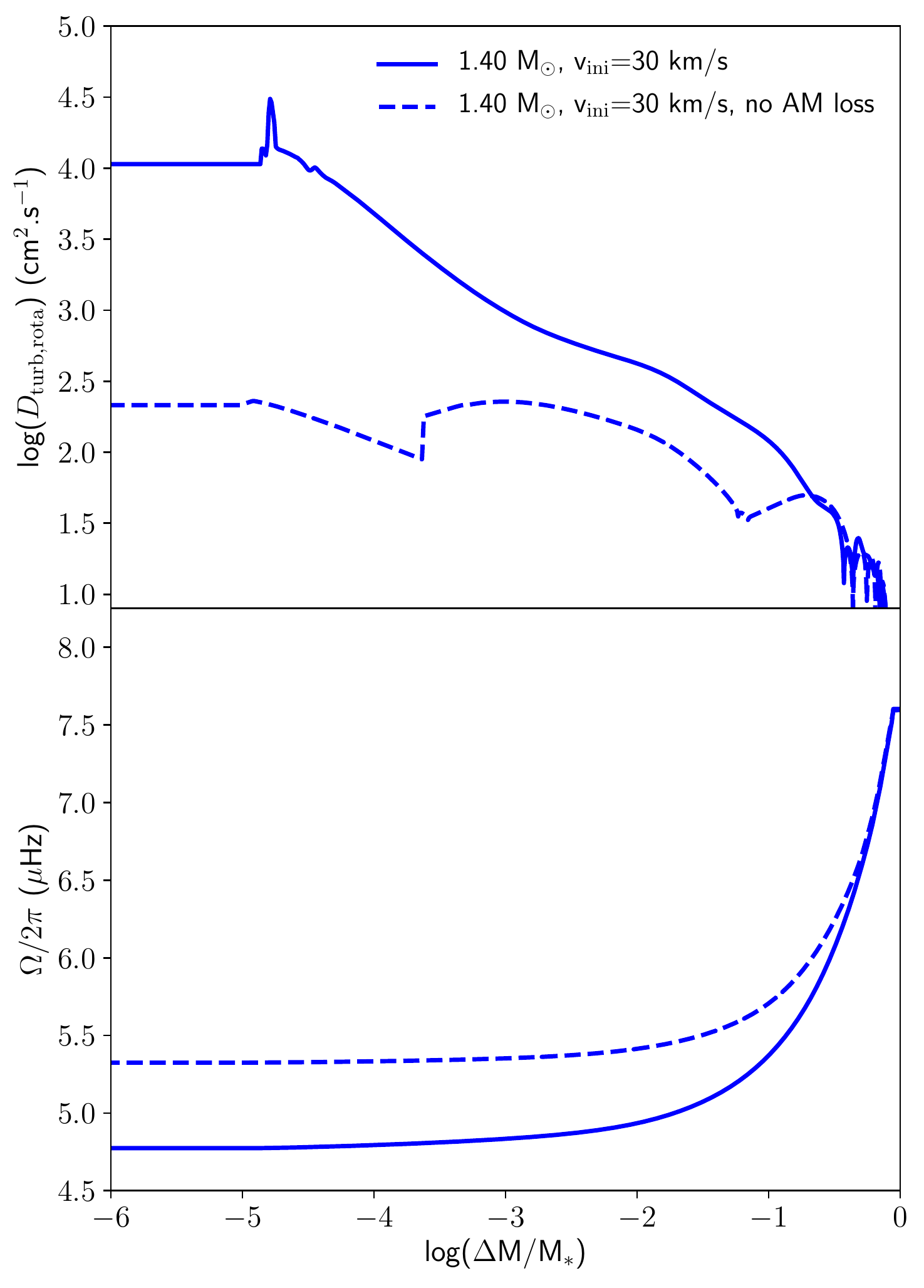}
\caption{Turbulent diffusion coefficient induced by rotation (upper panel) and rotation profile (lower panel) as a function of $\log (\Delta M/M^\ast)$ ($\Delta M$ is the mass above the considered region and $M^\ast$ is the mass of the star) at 1 Gyr for 1.40~M$_{\odot}$ models. The solid line corresponds to model B1 with magnetic braking whereas the dashed line is for model B5 without magnetic braking. The surface of the star is on the left of the figure.}
\label{fig:2}
\end{figure}

\subsubsection{Atomic diffusion and magnetic braking}\label{MB}

As discussed in Section \ref{coupling}, atomic diffusion and rotation are strongly coupled and this is mainly due to the magnetic braking. 

\medskip
\noindent {\bf Impact of magnetic braking on mixing:}
The main process  controlling the efficiency of the rotationally induced mixing  is the extraction of angular momentum at the surface. Without it, the mixing would be rather small. Figure \ref{fig:2} shows the profile of ${D_\mathrm{turb,rota}}$ for models B1 and B5. When there is a loss of angular momentum at the surface, the mixing is 1.5 orders of magnitude larger at the bottom of the surface convective zone (solid curve). This implies a stronger reduction of the efficiency of atomic diffusion. When the extraction is negligible, atomic diffusion  dominates. So the way we model the transport of angular momentum and its extraction is a key point for the transport of chemical elements.

\begin{figure}
\center
\includegraphics[scale=0.5]{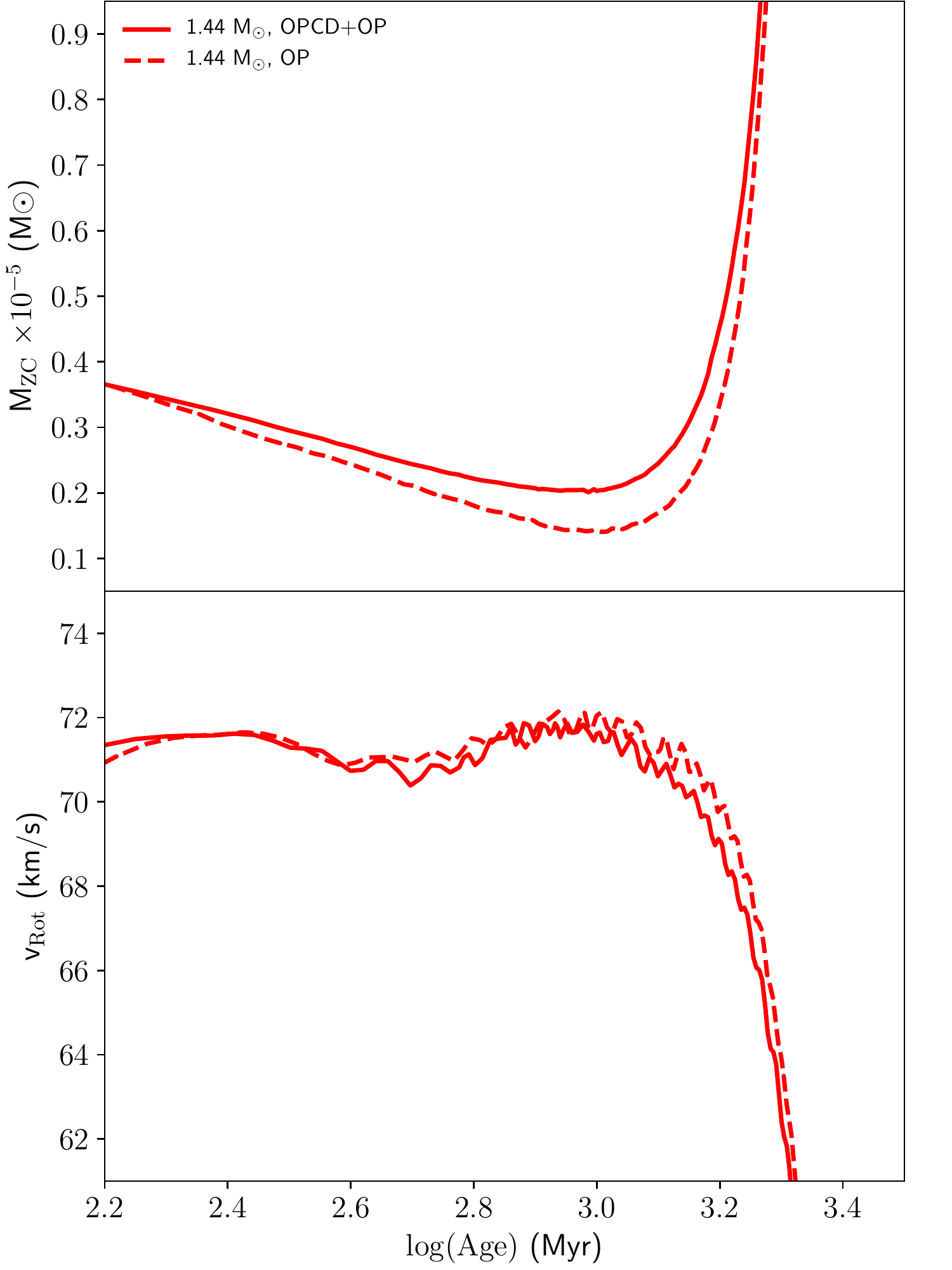}
\includegraphics[scale=0.47]{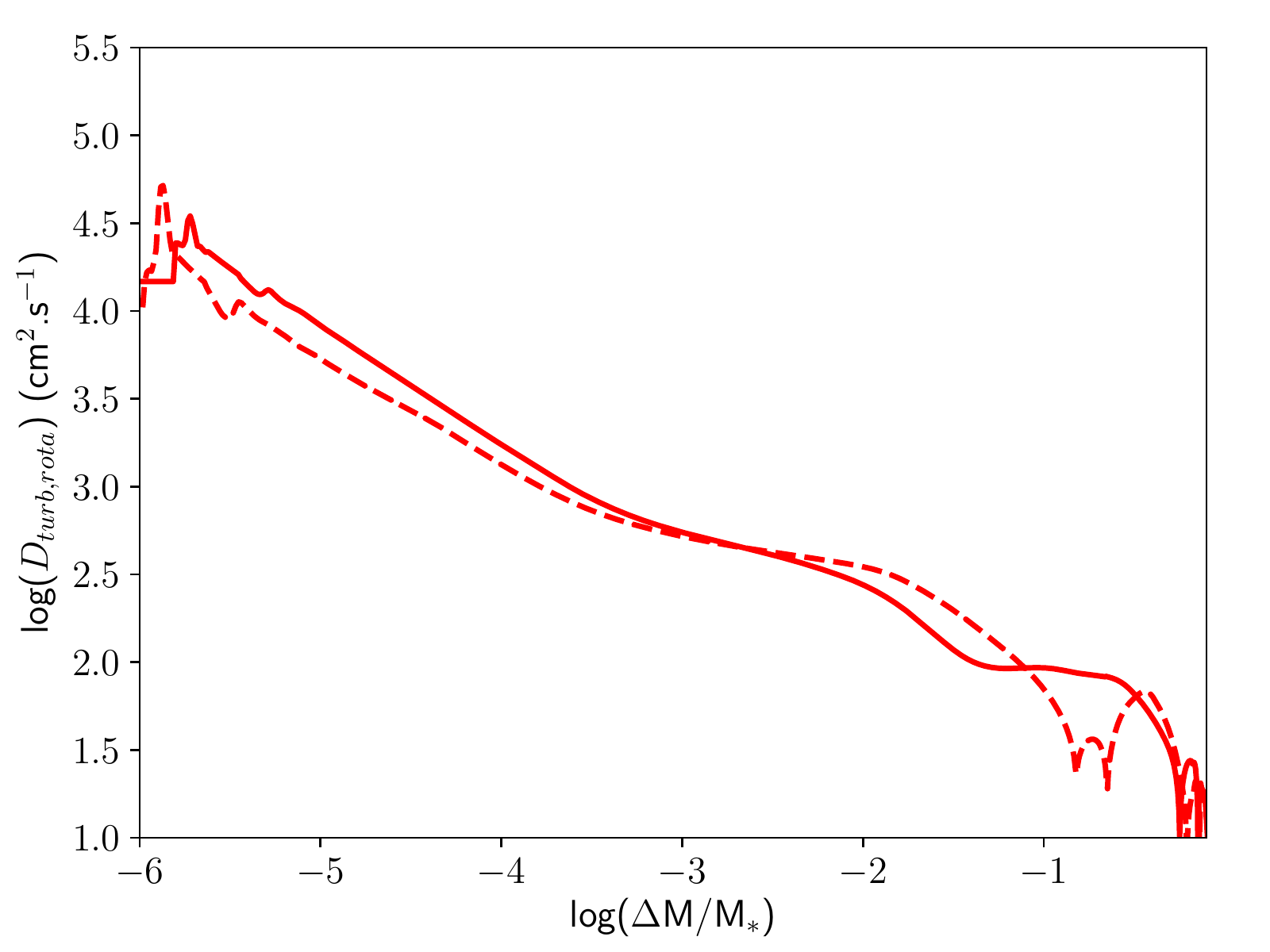}
\caption{Evolution of the mass of the surface convective zone (top panel) and of the surface rotation speed (middle panel) of 1.44~M$_{\odot}$ models (C2 and C5). In the C5 model, the opacities are calculated from an OP opacity table assuming a fixed, solar, mixture of chemical elements (dashed curves) while in the C2 model the opacities are calculated for the actual composition with the OPCD package. The lower panel shows the $D_\mathrm{turb, rota}$ profile for both models at 1 Gyr.}
\label{fig:3}
\end{figure}

\medskip
\noindent
{\bf Impact of atomic diffusion on magnetic braking:}
The extraction of angular momentum through magnetized winds depends on the ability of the star to maintain a magnetic field thanks to the dynamo process. This mainly depends on the size of the convective envelope and on processes that may affect it. The thicker the envelope is, the more efficient the braking of the star is. In particular, the size of the convective zone can be affected by atomic diffusion when heavy elements are accumulated and induce a local increase of the opacity \citep[e.g.][]{turcotte98b,deal18}. To illustrate this effect on opacity, we computed a model (C5), which is the same as C2, but in which we did not recompute the Rosseland mean opacity at each mesh point and each time step when the mixture is affected by diffusion. In this model, because of the accumulation of heavy elements at the surface (especially iron in this case) the increase of opacity at the bottom of the surface convective zone is strongly underestimated. This explains why the size of the surface convective zone is smaller in model C5 than in model C2 as shown in the top panel of Fig. \ref{fig:3}. The maximum relative difference in the convective zone mass reaches 45\% .

As the convective zone is larger when accounting properly for the abundance variations and opacities, the braking due to magnetized wind is then more efficient. Fig. \ref{fig:3} (middle panel) shows the evolution of the rotation speed with time. The rotation speed of the C2 model is lower than the one of the C5 model. The difference reaches a maximum of 1~km/s. The enhanced efficiency of the extraction of angular momentum in model C2 leads to a mixing induced by the rotation 40\% larger than in  model C5 (between $\log (\Delta M/M^\ast)= -6~\mathrm{and}~-4$). As the mixing is less efficient in C5, the accumulation of iron in the surface convective zone is then larger.

This example is a good illustration of the coupling that exists between transport processes. In this case the coupling occurs through opacity, due to the variation of the chemical composition with time. This confirms that it is no longer possible to use opacity tables with a fixed mixture of chemical elements for this type of star when mixing processes affecting elements individually are taken into account. Computation of stellar models then requires opacity tables such as provided by the publicly available OPCD package, for example.

\subsection{Uniform versus differential rotation}\label{solid}

The theory of the transport of angular momentum and associated transport of chemicals in stars is still far from being understood. The observations show that the current theory underestimates the transport of angular momentum between the core and the surface of stars, low and intermediate mass main sequence stars tending to rotate nearly uniformly in their radiative regions  \citep[e.g][]{deheuvels12,aerts18,ouazzani19,buldgen19}.

Accordingly in this section, we study the impact on the surface abundances of a uniform rotation maintained along the main sequence evolution. We build stellar models assuming solid body rotation under the assumption that the angular momentum transport is efficient enough to maintain a uniform rotation despite structural changes and hydrodynamical instabilities.We further assume that this extra transport of angular momentum does not affect the  mixing of chemical elements. In this framework the modelling assumes neither shear-induced transport nor meridional circulation. The characteristics of the models (A6, B6 and C6) are listed in Table \ref{table:2}. We use a value of $8.3\times10^{30}$~erg for the constant of the extraction of angular momentum in models with uniform rotation taking into account \cite{matt19} recommendations.

\begin{table}
\centering
\caption{Inputs of the models with uniform rotation} 
\label{table:2}
\begin{tabular}{l|c|c|c|c}
\noalign{\smallskip}\hline\hline\noalign{\smallskip}
Model & A6 & B6$_1$ & B6$_2$ & C6\\
\noalign{\smallskip}\hline\noalign{\smallskip}
Mass (M$_{\odot}$) & 1.30  & \multicolumn{2}{c|}{1.40} & 1.44\\ 
\noalign{\smallskip}\hline\noalign{\smallskip}
AM Loss & \multicolumn{4}{c}{Matt19}\\
\noalign{\smallskip}\hline\noalign{\smallskip}
v$_{ini}$ (km/s) & 30 & 30 & 80 & 30\\
\noalign{\smallskip}\hline\noalign{\smallskip}
Opacities & \multicolumn{3}{c}{OPCD+OP} \\
X$_{ini}$ & \multicolumn{3}{c}{0.7280}  \\ 
Y$_{ini}$ & \multicolumn{3}{c}{0.2578} \\
Z/X$_{ini}$ & \multicolumn{3}{c}{0.0195} \\
$\alpha_\mathrm{CGM}$ & \multicolumn{3}{c}{0.68}  \\ 
Mixture & \multicolumn{3}{c}{AGSS09}  \\ 
EoS & \multicolumn{3}{c}{OPAL2005}  \\
Nuc. react. & \multicolumn{3}{c}{NACRE+LUNA} \\ 
Core ov. & \multicolumn{3}{c}{0.15} \\    
\noalign{\smallskip}\hline\noalign{\smallskip}
\end{tabular}
\end{table}

\begin{figure}
\center
\includegraphics[scale=0.5]{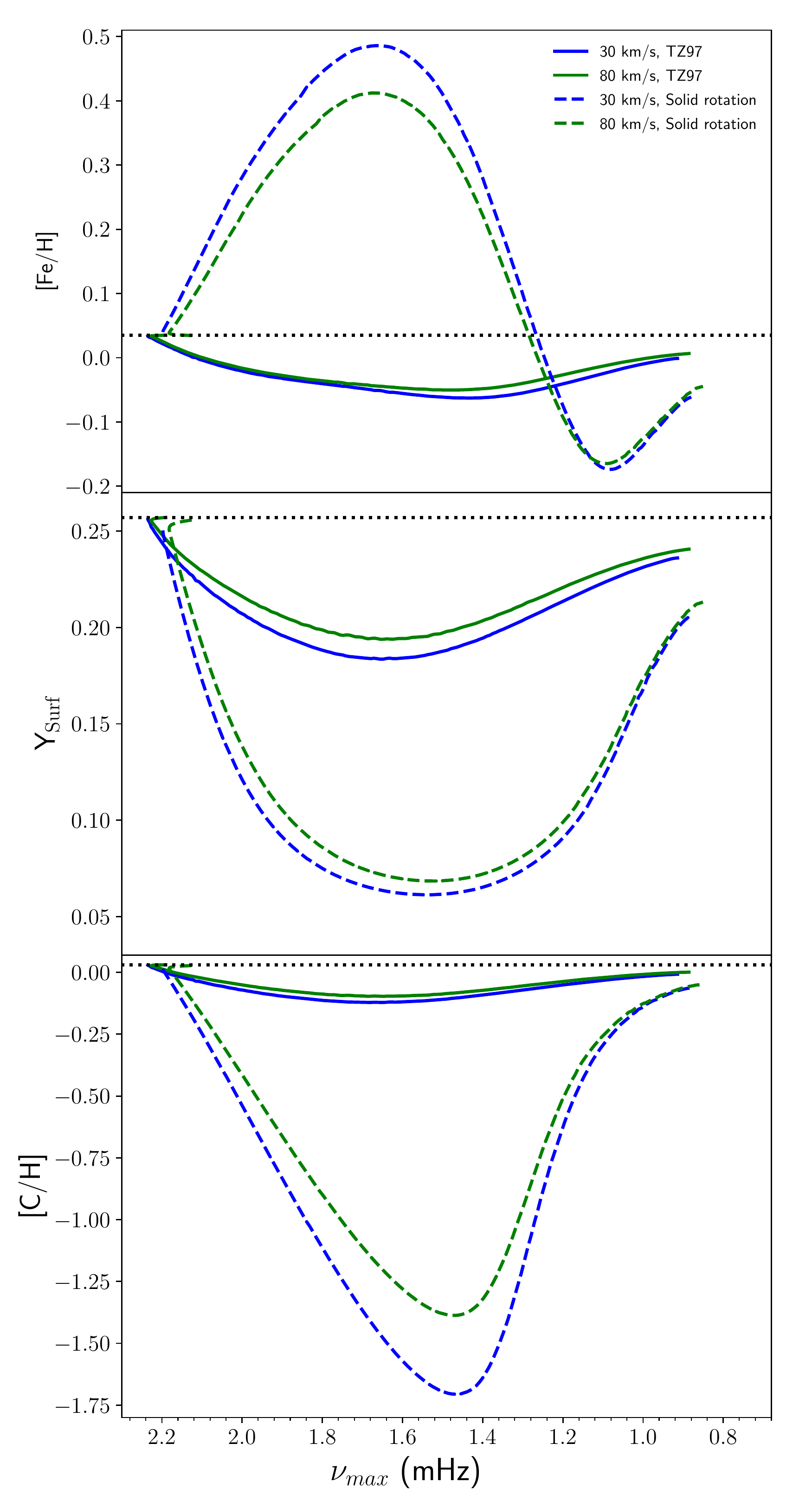}
\caption{Evolution of [Fe/H] in dex (top panel), helium $Y_\mathrm{Surf}$ in mass fraction (middle panel) and the surface [C/H] in dex (lower panel) as a function of $\nu_\mathrm{max}$ for 1.4~M$_{\odot}$ models with differential (solid curves) and uniform (dashed curves) rotation at two rotation speeds. Horizontal dotted lines represent the initial abundance.}
\label{fig:4}
\end{figure}

Figure \ref{fig:4} shows the surface abundances of iron, helium and carbon for the 1.4 M$_\odot$ models. As there is no transport of chemical elements induced by rotation, the only impact on the transport is due to the inflation of the star due to rotation. As a consequence, the effects of atomic diffusion are not reduced and the surface abundances of helium and carbon are not consistent with observations \citep[e.g.][]{ghazaryan18,verma19}. The increase of the initial rotation speed helps to reduce the impact of atomic diffusion but an initial speed of 80~km/s is still not sufficient (green curves of Fig. \ref{fig:4}.

We also investigated the case when  a uniform rotation is induced by an unknown additional transport of angular momentum. We do so as in \cite{eggenberger19} (and references therein), by including an additional  viscosity, $\nu_{V,\mathrm{add}}=10^8$ cm$^2$.s$^{-1}$, (i.e. $D_v+\nu_{V,\mathrm{add}}$ in eq. \ref{Dv}). The value is  calibrated by \cite{ouazzani19} on $\gamma$-Doradus stars. We assume that this additional transport of angular momentum does not impact directly the transport of chemical elements. In this case the shear-induced transport is negligible and the only transport of chemical element comes from the interaction between meridional circulation and horizontal shear turbulence. This interaction results in a process which behaves like a diffusive process with a coefficient given by $D_\mathrm{eff}=(r U_2)^2/30 D_h$. In our case $D_h$ is very large compared with $(r U_2)^2$ and $D_\mathrm{eff}$ is therefore small. The transport of chemical elements is this negligible. The resulting surface abundances are similar to the above case of imposing directly a uniform rotation. 

From there on, we conclude that in the framework of a forced solid body rotation with no impact on rotational mixing, only chemically peculiar stars (Fm stars in the 1.4~M$_\odot$ and 1.44~M$_\odot$ cases) should exist and be observed. However, this scenario is not supported by observations of most of the \textit{Kepler} stars \citep{bruntt12,brewer16}. This indicates that, an extra transport process of chemical elements is also missing, likely associated with the extra angular momentum transport needed to account for observations.

\section{Inferring stellar parameters from observational constraints}

Our goal in this section is to estimate the impact of neglecting rotation and atomic diffusion in stellar models on the inference of stellar parameters. For that purpose, we compare the `true' stellar parameters of artificial stars with those obtained by the AIMS optimization code \citep{lund18,rendle19} using a grid of models in which these ingredients are lacking. The artificial stars are represented by the models previously discussed, (A1, A3, B1, B3, C1, C2, C3, C4) in which transport by rotation and diffusion are taken into account, the models of 1.3 and 1.4~M$_{\odot}$ including radiative accelerations of \cite{deal18}, and models of 1.3 and 1.4~M$_{\odot}$ including the same physics as the grid for tests purpose. We first describe the methodology.

\subsection{The AIMS optimization code}

The AIMS optimization code applies a Monte Carlo Markov Chain (MCMC) approach in order to find a representative sample of stellar models which fit a given set of classic and seismic constraints. This sample is subsequently used to find the best-fitting values, error bars, and posterior probability distribution functions (PDFs) for the different stellar properties. In order to gain computation time, the AIMS code uses a precomputed grid of stellar models which includes global stellar properties such as mass and age as well as pulsation spectra, and interpolates within this grid for each MCMC iteration. Interpolation is carried using a multi-dimensional Delaunay tessellation between evolutionary tracks and a simple linear interpolation along evolutionary tracks. For the seismic constraints, AIMS will treat the pulsation frequencies as independent normally distributed random variables. During the fitting process, it is possible to fit frequency combinations such as frequency ratios rather than individual frequencies -- AIMS will automatically calculate the resultant correlation matrix for such combinations. For the classic constraints, a larger variety of probability distribution functions can be applied. In what follows, we used normal distributions, truncated at one $\sigma$, in order to avoid having the optimization method go into the wings of the probability distribution.

\subsection{Input to AIMS: Grids of stellar models and associated oscillation frequencies}

When using the AIMS code, we consider grids of stellar models that do not include any transport processes but  convection. The other input physics are kept the same as for the artificial stars (Table \ref{table:1}). The stellar models of the grids are also computed with the CESTAM code. The oscillation frequencies are computed using ADIPLS \citep{christensen08} for l=[0,3] and, n=[0,30]. The characteristics of Grid 1, (hereafter the standard physics grid) are presented in Table \ref{table:3}.

\begin{table*}[ht]
\centering
\caption{\label{table:3} Parameters of the grids used in the AIMS stellar parameter optimization pipeline.}

\begin{tabular}{ccc|ccc}
\noalign{\smallskip}\hline\hline\noalign{\smallskip}
	Grid 1&  &  & Grid 2&  &\\
\noalign{\smallskip}\hline\noalign{\smallskip}
	Variables & Range & Steps & Variables & Range & Steps \\
\noalign{\smallskip}\hline\noalign{\smallskip}
	M (M$_\odot$) & 1.2-1.625 & 0.025 & M (M$_\odot$) & 1.2-1.5 & 0.025\\
	Y$_\mathrm{ini}$ & 0.15-0.28 & 0.01 & Y$_\mathrm{ini}$ & 0.24-0.28 & 0.0025 \\
	Z$_\mathrm{ini}$ & 0.006-0.02 & 0.002 & Z$_\mathrm{ini}$ & 0.010-0.018 & 0.001 \\
\noalign{\smallskip}\hline\noalign{\smallskip}
	l & 0-3 & 1 & l & 0-3 & 1 \\
	n & 0-30 & 1 & n & 0-30 & 1 \\
\noalign{\smallskip}\hline\noalign{\smallskip}
\end{tabular}
\end{table*}

\subsection{Input to AIMS: the "observational" constraints of the artificial stars}

We used the usual classical constraints: $T_\mathrm{eff}$, luminosity and [Fe/H] and included 
for the seismic constraints the individual oscillation frequencies with degree $l=0$ to 2. We computed the frequency at maximum power $\nu_\mathrm{max}$ using a scaling relation \citep{kjeldsen95} and chose ten frequencies above and below $\nu_\mathrm{max}$. The uncertainties on the frequencies were set to $0.5~\mu\mathrm{Hz}$ and for the classical constraints respectively $70~K$ ($T_\mathrm{eff}$), $0.05$ (luminosity) and $0.1~\mathrm{dex}$ ([Fe/H]).

We ended up with $\approx$ 63 individual frequencies and three classical parameters for the "observational" constraints. We also  carried out inference optimization  using frequency ratios rather than individual frequencies. In that case, we also added $\Delta\nu_\mathrm{moy}$ (the mean large frequency separation) and the three lowest frequencies of $l=0$ to the set of constraints. The inclusion of low individual frequencies allows us to constrain better the mass (Serenelli, A, priv. comm.) while not being too sensitive to surface effects. The number of frequencies is quite large because our aim is to assess how large are the systematic errors on the stellar parameter determination when the physical description of chemicals transport is not properly modeled. We nevertheless also tested cases in which the $l=2$ frequencies were removed, or reducing the frequency uncertainties to $0.1~\mu\mathrm{Hz}$. The results were identical.

\subsection{Testing the accuracy of the AIMS code}

AIMS is a powerful tool to infer stellar parameters but its precision depends on the stellar model grid it uses \citep{rendle19}. In order to determine the uncertainties using Grid 1, we chose as pseudo target stars, models with the physics of Grid 1 but which do not belong to the grid. The adopted models for the pseudo target stars are listed in Table \ref{table:4}.

\begin{table}[ht]
\centering
\caption{\label{table:4} Parameters of the test models.}

\begin{tabular}{c|c|c}
\noalign{\smallskip}\hline\hline\noalign{\smallskip}
	Models& A7, A8, A9 & B7, B8, B9 \\
\noalign{\smallskip}\hline\noalign{\smallskip}
	M (M$_\odot$) & 1.3  & 1.4\\
	Y$_\mathrm{ini}$ & 0.2578 & 0.2578\\
	Z$_\mathrm{ini}$ & 0.0142 & 0.0142\\
	X$_\mathrm{C}$ & 0.6-0.4-0.2 & 0.6-0.4-0.2\\
\noalign{\smallskip}\hline\noalign{\smallskip}
\end{tabular}
\end{table}

\begin{table*}[ht]
\centering
\caption{\label{table:5} Results of the optimization for the test models. "I" stands for optimization with individual frequencies and "R" stands for optimization with frequency ratios.}
\begin{tabular}{c|ccc|ccc|ccc|cc}
\noalign{\smallskip}\hline\hline\noalign{\smallskip}
    Models & \multicolumn{3}{c|}{A7} & \multicolumn{3}{c|}{A8}& \multicolumn{3}{c|}{A9} & \multicolumn{2}{c}{Error max. (\%)}\\
\noalign{\smallskip}\hline\noalign{\smallskip}
	 & Inputs & I & R & Inputs& I & R & Inputs& I & R & I & R\\
\noalign{\smallskip}\hline\noalign{\smallskip}
	M (M$_\odot$)    & 1.300  & 1.304  & 1.303  & 1.300  & 1.330  & 1.418  & 1.300  & 1.306  & 1.308  & 2.3 & 9\\
	Age (Gyr)        & 1.046  & 1.063  & 1.060  & 2.346  & 2.314  & 2.577  & 3.326  & 3.351  & 3.350  & 1.6 & 10\\
	R (R$_\odot$)    & 1.303  & 1.305  & 1.304  & 1.454  & 1.470  & 1.499  & 1.638  & 1.640  & 1.641  & 1.1 & 3.1 \\
	Y$_\mathrm{ini}$ & 0.2578 & 0.2541 & 0.2546 & 0.2578 & 0.2500 & 0.1800 & 0.2578 & 0.2530 & 0.2528  \\
	Z$_\mathrm{ini}$ & 0.0142 & 0.0140 & 0.0140 & 0.0142 & 0.0156 & 0.0131 & 0.0142 & 0.0140 & 0.141 \\
	X$_\mathrm{C}$   & 0.60   & 0.60   & 0.60   & 0.40   & 0.42   & 0.46   & 0.20   & 0.20   & 0.20\\	
	$\bar{\rho}$     & 0.827  & 0.827  & 0.827  & 0.596  & 0.591  & 0.593  & 0.417  & 0.417  & 0.417\\
	$\log g$         & 4.322  & 4.322  & 4.322  & 4.227  & 4.227  & 4.234  & 4.123  & 4.124  & 4.124\\
	
\noalign{\smallskip}\hline\noalign{\smallskip}
 & \multicolumn{3}{c|}{B7} & \multicolumn{3}{c|}{B8}& \multicolumn{3}{c|}{B9}\\
\noalign{\smallskip}\hline\noalign{\smallskip}
	M (M$_\odot$)    & 1.400  & 1.441  & 1.424  & 1.400  & 1.400  & 1.400  & 1.400  & 1.407  & 1.409  & 2.9 & 1.7 \\
	Age (Gyr)        & 0.866  & 0.806  & 0.993  & 1.926  & 1.942  & 1.941  & 2.686  & 2.708  & 2.688  & 7.5 & 23\\
	R (R$_\odot$)    & 1.427  & 1.441  & 1.434  & 1.617  & 1.617  & 1.617  & 1.844  & 1.847  & 1.849  & 1 & 0.5\\
	Y$_\mathrm{ini}$ & 0.2578 & 0.2368 & 0.2263 & 0.2578 & 0.2560 & 0.2561 & 0.2578 & 0.2528 & 0.2562\\
	Z$_\mathrm{ini}$ & 0.0142 & 0.0140 & 0.0120 & 0.0142 & 0.0140 & 0.0140 & 0.0142 & 0.0141 & 0.141\\
	X$_\mathrm{C}$   & 0.60   & 0.63   & 0.62   & 0.40   & 0.40   & 0.40   & 0.20   & 0.20   & 0.21\\	
	$\bar{\rho}$     & 0.679  & 0.678  & 0.680  & 0.466  & 0.466  & 0.466  & 0.314  & 0.315  & 0.314\\
	$\log g$         & 4.275  & 4.279  & 4.278  & 4.166  & 4.166  & 4.166  & 4.052  & 4.053  & 4.053\\
\noalign{\smallskip}\hline\noalign{\smallskip}
\end{tabular}
\end{table*}

The comparison between the parameters of the input models and the determination with AIMS using Grid 1 are presented in Table \ref{table:5}. We can see that one obtains a precision of the order of 2.9\% for the mass, 7.5\% for the age and 1\% for the radius when using individual frequencies. In most of these six cases, AIMS finds the exact value; the most uncertain parameter is the helium content. Adding the helium content as an observational constraint is definitely helpful in lifting the degeneracy (Mass/Y). For example, for the A8 model, the mass inference goes from 1.33 to 1.325~M$_\odot$ (using grid 1) when helium is added as observational constraint. The mass inference is 1.32~M$_\odot$ when using grid 2 (denser grid, see Table \ref{table:5}) and goes to 1.29~M$_\odot$ taking into account the helium constraint.

The differences in age, mass and radius are also shown in Fig. \ref{fig:5}. This figure represents the departures of the values inferred by AIMS from the values of the input models. Two-$\sigma$ uncertainties are adopted which are obtained from AIMS probability distributions for each parameter. Note that when the   uncertainties appear unreasonably small, it only means that the number of accepted models in the grid, taking into account the "observational" constraints is small. This signals (except for model A8) that the physics of the grid is too different from the input models to find a reliable solution. This is mainly due to the [Fe/H] constraint, as the chemical composition is the main difference between the grid and inputs models. For model A8, the small uncertainties cannot come from a different physical description, by construction; they are due to a small number of accepted models. The results slightly improves when using grid 2. We conjecture that for this model, the problem comes from the density of the grid. 

We performed the same comparison using a second set of constraints i.e. frequency ratios, mean $\Delta\nu$ and the three lowest $l=0$ frequencies instead of individual frequencies as seismic constraints. The precision is slightly worse compared to when using individual frequencies. Ratios then appear less effective than individual frequencies. This must be attributed to the fact that in our theoretical study with individual frequencies, our set of frequencies is quite large with many low frequencies (best suited to constrain the structure). In such ideal cases (where no surface effects pollute the frequencies), the optimization using individual frequencies is quite efficient at finding an optimal model. Reducing by a factor of two the number of frequencies provided results with individual frequencies equivalent to those obtained with frequency ratios. We then decided to continue this study, using only the large set of individual frequencies as seismic constraints.

The same tests were performed using Grid 2 presented in Table \ref{table:3}. This grid covers the same range of mass but a smaller range of chemical compositions. It is finer and the time steps between models was divided by a factor of two. The maximum errors on age using individual frequencies and Grid 1 was 7.5\% for the D model. With this grid the error drops to 1.6\%. All the other determinations are slightly improved, with error reductions of a few 0.1\%, but the worst cases are drastically improved. The error on the parameter determinations comes mainly from the quality of the grid but the method is robust. Computing a new grid with the resolution of Grid 2 but the range of parameters of Grid 1 would have required to much computation resources and would not have change the conclusion of this paper. This is the reason why we performed the study with Grid 1.

\subsection{The solar case}
As a further test, we consider the solar case. We studied  on one hand a model close to the solar model with parameters included in the grid and on the other hand a set of noisy frequencies -taken from \cite{silva17}- which were determined from solar data but degraded to the level of quality corresponding to the Kepler observations . We computed a new grid which includes the same physics as Grid 1 and 2 but takes into account atomic diffusion (without radiative accelerations). The properties of the grid are [0.9; 1.1; 0.025]M$_\odot$ for the mass, [0.22; 0.27; 0.01] for $Y_\mathrm{ini}$ , [0.010; 0.018; 0.002] for $Z_\mathrm{ini}$ and [0.6; 0.8; 0.02] for $\alpha_\mathrm{CGM}$. No core overshoot is taken into account. We chose the two term surface correction expression from \cite{ball14} to account for surface effects in the real data. 

The results obtained using AIMS (2-$\sigma$ errors) for  the model are a mass of $0.997 \pm 0.011$~M$_\odot$ (model mass: 1.0~M$_\odot$), a radius of $0.998 \pm 0.007$~R$_\odot$ (model radius: 0.999~R$_\odot$) and an age of $4.672 \pm 0.400$ Gyr (model age: 4.606~Gyr). The degraded solar data gave us a mass of $0.987 \pm 0.014$~M$_\odot$, a radius of $0.994 \pm 0.042$~R$_\odot$ and an age of $4.845 \pm 0.140$ Gyr. In both cases, the method and the grid give results very close to real values of the Sun parameters. 

\section{Impact of neglecting transport processes on stellar parameter inference}\label{resultAIMS}

In this section, we assess the impact of neglecting atomic diffusion (including radiative accelerations) and/or rotationally induced mixing on the seismic inference of stellar parameters (mass, radius, age, chemical composition). 

\subsection{Models with atomic diffusion and without rotation}\label{AIMSgrad}

\begin{figure*}[ht]
\center
\includegraphics[scale=0.7]{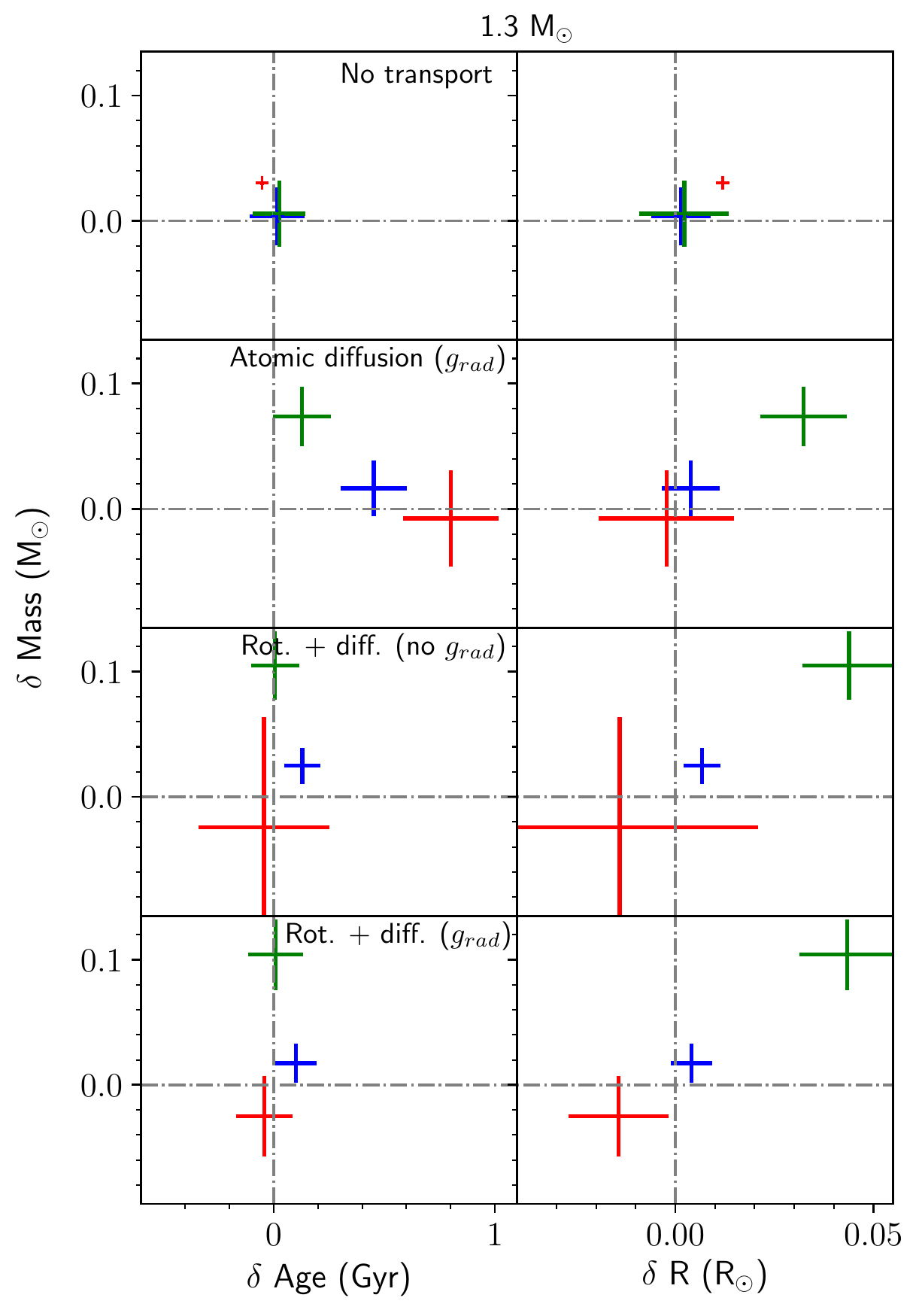}
\includegraphics[scale=0.7]{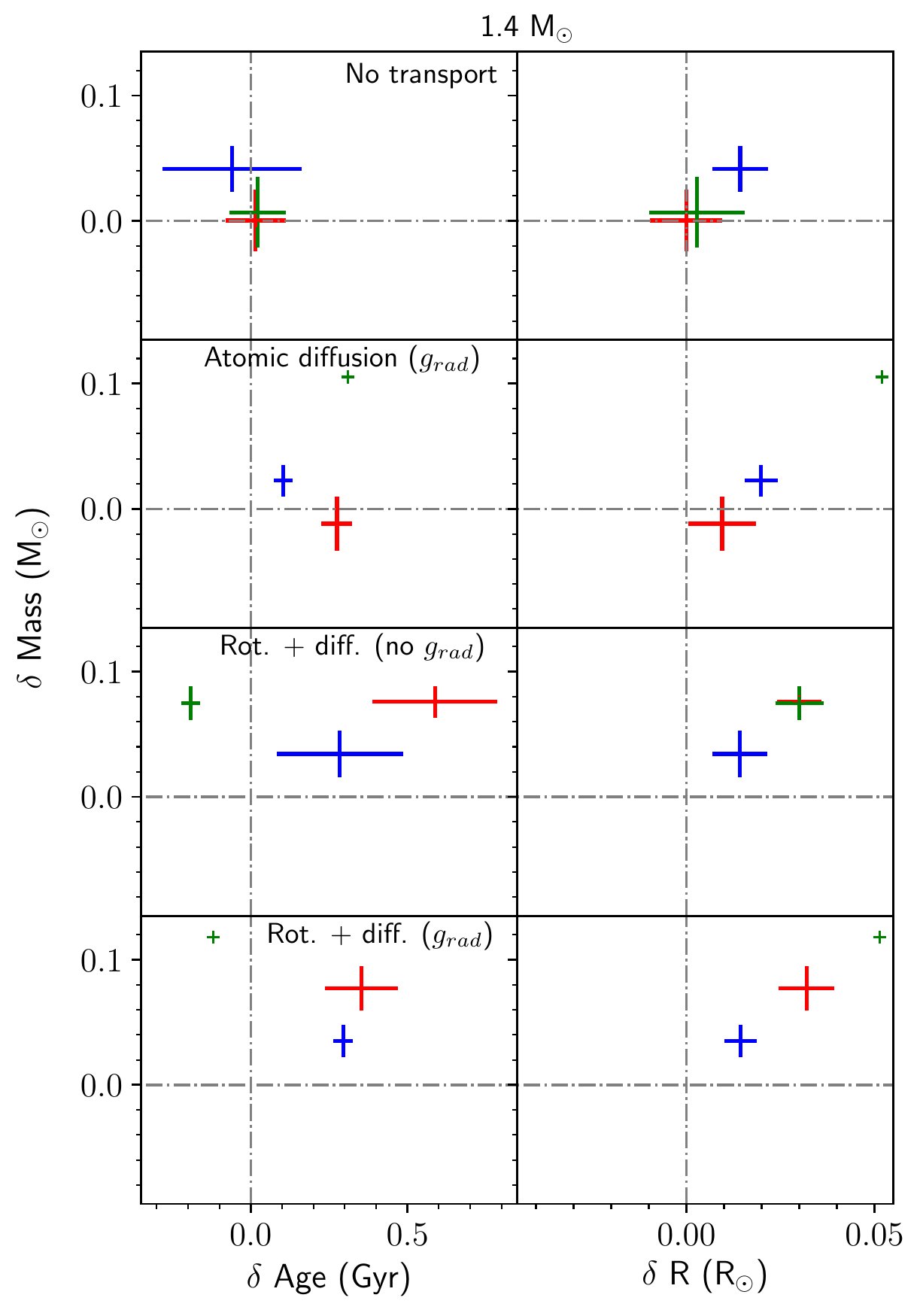}
\caption{Deviation between the real value of mass, age and radius, and the inferred value by AIMS (using Grid 1). The left panel shows the results for the 1.3~M$_\odot$ model and the right panel shows the results for the 1.4~M$_\odot$ model. The crosses (blue: $X_c=0.6$, red: $X_c=0.4$, green, $X_c=0.2$) represent the deviations with 2-$\sigma$ uncertainties from AIMS probability distributions. }
\label{fig:5}
\end{figure*}

\begin{figure}[ht]
\center
\includegraphics[scale=0.7]{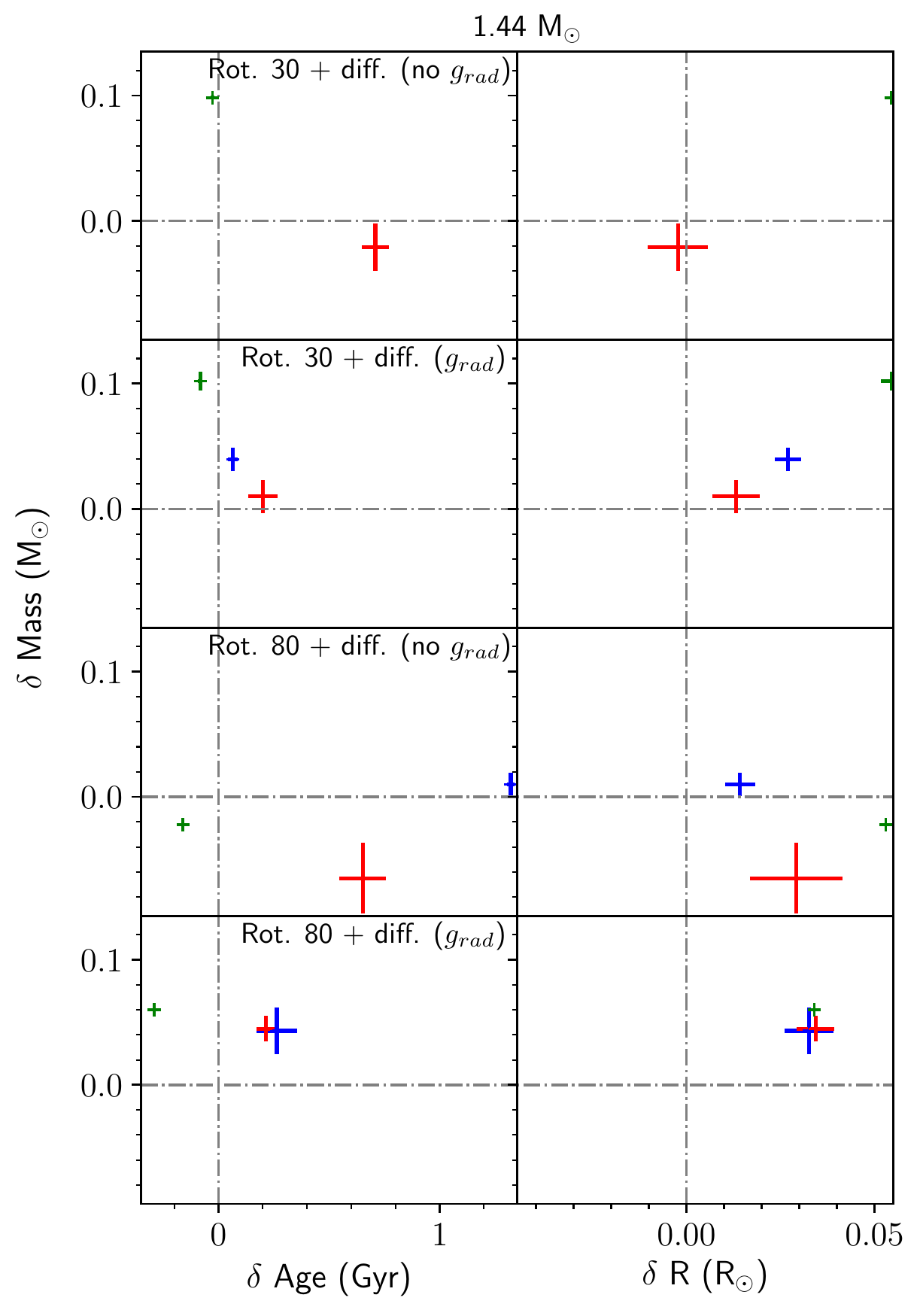}
\caption{Deviation between the real value of mass, age and radius, and the inferred value by AIMS (using Grid 1) for 1.44~M$_\odot$ models. The legend is the same as Fig. \ref{fig:5} }
\label{fig:6}
\end{figure}

We first tested the impact of atomic diffusion (including radiative accelerations) on the stellar parameter determinations. To this end, we used the 1.3 and 1.4~M$_\odot$ models at solar metallicity presented in \cite{deal18} at three epochs on the main sequence. These models include atomic diffusion unlike our stellar model grid. They do not include rotationally induced mixing. The results from the optimization are presented Fig. \ref{fig:5}. The same trend with evolution seems to appear for 1.3 and 1.4~M$_\odot$ models. The deviation from the true age increases up to $X_c=0.4$ and then decreases at $X_c=0.2$ or at least is not larger for the 1.4~M$_\odot$ model. The cause for this deviation arises because of a difference in the surface iron content, [Fe/H]. Indeed since we considered a grid without transport of chemical elements in radiative regions, the models of the grid therefore keep the same [Fe/H] all over the main-sequence. This is not the case for the models representing the artificial stars for which we want to determine the stellar parameters because they include atomic diffusion. In order to match the value of the [Fe/H] of the artificial star at its evolutionary stage, the optimization method then searches for a model with a different [Fe/H] than the birth one. Then the optimization process will infer an incorrect set of stellar parameters, mainly a wrong age, due to this metallicity effect. The deviation is smaller at the end of the main sequence because the surface convective zone starts deepening and erasing the abundance variations created during the evolution. The impact on the age is larger when the "observed" models have a surface [Fe/H] lower than the initial one (1.3~M$_\odot$ models, low effect of radiative accelerations on iron) rather than the opposite (1.4~M$_\odot$ models, large effect of radiative accelerations on iron).

The deviations for the radius and mass at $X_c=0.6$ and $0.4$ are very small and increase at the end of the main-sequence. As the star evolves, atomic diffusion has more and more time to act. One of the effects of radiative accelerations in particular, is to reduce the density in the affected regions. This causes a more significant increase of the radius at the end of the main sequence. For a given $\Delta \nu$ (i.e a given mean density), a larger radius translates directly into a larger inferred mass.

The deviations are most of the time below 5\% and 2\% respectively for the mass and radius, but the age is strongly impacted (above 10\%). These results are upper limit deviations because atomic diffusion is in competition with other transport processes. In any case, it is important to keep in mind that this process occurs anyway, and can lead to significant errors in parameter inference when neglected. 

\subsection{Models with rotation and atomic diffusion}

We now turn to the case where some competition exists between atomic diffusion and rotationally induced mixing. 

\subsubsection{Without radiative accelerations}

We first test the impact of taking rotation and diffusion without radiative accelerations into account in the models, on stellar parameter inference (models A3, B3, C3 and C4 of Table \ref{table:1}). The initial rotation speed is 30 km/s for the 1.3 and 1.4~M$_\odot$ models, and both 30 and 80 km/s for the 1.44~M$_\odot$ models. The deviation of AIMS inferences from the ``real'' values are presented Fig. \ref{fig:5}.

For the 1.3~M$_\odot$ models, the age deviation remains small  over the entire main sequence. The important mixing induced by rotation limits the variations of the surface [Fe/H] due to atomic diffusion, which therefore remains small. At a given evolutionary state, rotation induces an effective gravity which is smaller than the one of an equivalent star without rotation. At a constant mass, it induces a larger radius. At a given mean $\Delta\nu$, the optimisation will then infer a larger mass.

For the 1.4~M$_\odot$ models, the impact of atomic diffusion has become strong while the efficiency of the extraction of angular momentum has decreased enough so that the surface abundances significantly vary compared with their initial values. This leads to inferred age values which significantly deviate from the ``real'' value. The maximum deviation occurs for $X_c=0.4$ when the [Fe/H] is more than 0.1~dex lower than the initial value. The deviation of the radius is smaller than for the 1.3~M$_\odot$ models, hence leading to a smaller deviation in mass.

The 1.44~M$_\odot$ models (Fig. \ref{fig:6}) show even larger age deviations than the 1.4~M$_\odot$ models due to larger [Fe/H] variations during the main sequence. But radii and masses are quite well recovered by AIMS despite the use of the standard physics grid. The impact of the initial rotation speed is mainly visible on the radius deviations, which increase with the rotation speed.

In a nutshell, the main impact of rotation/diffusion on the stellar parameter inference is upon the age, as expected. The impact on the age can be larger than 1~Gyr for stars with masses larger than 1.4~M$_\odot$ models. These examples show that for stars with masses larger than 1.3~M$_\odot$, both processes should be taken into account, in order to infer more accurate stellar parameters.

\subsubsection{With radiative accelerations}

When inference of stellar parameters are carried out for models including radiative accelerations i.e. models A1, B1, C1 and C2, the impact is smaller than the one due to rotation. For the 1.3~M$_\odot$ models, inferences are almost identical because radiative accelerations are not efficient (Fig. \ref{fig:5}). In the other cases (1.4 and 1.44~M$_\odot$ models), the scatter around the real values of age, mass and radius are smaller than the one due to rotation (see Fig. \ref{fig:5} and \ref{fig:6}). 

For the 1.4~M$_\odot$ models, the maximum difference in age is 0.6~Gyr and goes down to 0.35~Gyr when radiative acceleration are taken into account. At 1.4~M$_\odot$ it goes from 0.8 to 0.2 and from 1.15 to 0.3 Gyr respectively for models with a surface initial rotation speed of 30 and 80 km/s. This is because radiative accelerations reduce the [Fe/H] variations and make it closer to its initial value. The deviation from the standard models then remains small. For the mass and radius, the reduction of the deviation is smaller in both the 1.4 and 1.44~M$_\odot$ models.

As already mentioned, the transport of angular momentum is underestimated in the current rotating models \citep[e.g][]{deheuvels12,aerts18,ouazzani19,buldgen19}. When considering uniformly rotating models equivalent to the ones presented in Section \ref{solid}, atomic diffusion dominates over the mixing induced by rotation and the impact on the age is  more significant (close to the deviations presented in Section \ref{AIMSgrad}). Hence like the transport of chemical elements, the transport of angular momentum is a key to build models with accurate stellar parameters for solar-like oscillating stars.

\subsection{Impact of core overshoot on the age}

The uncertainties on ages due to rotation and atomic diffusion can also be compared with the ones arising from uncertainties on the core overshoot extension. We computed models with the same input physics as model A1, but without rotation. We compute the opacities with the OP opacity table instead of the OPCD package and  tested that it has no impact on the age at the end of the main sequence. We assumed several values for the extension of  the core overshoot. We determined that in order to obtain the same age at the end of the main-sequence as model A1, with a model that does not take rotation into account, the core overshoot should be 0.18~$H_p$ (instead of 0.15~$H_p$). The age difference at the end of the main sequence between model A1 and the same model without rotation is 120~Myr ($\approx 3\%$ of the total age). A slight degeneracy appears here between the impact on the age of rotation and core overshoot.

\subsection{\textit{Kepler Legacy} stars}\label{kepler}

\begin{table*}[ht]
\centering
\caption{\label{table:8} Parameters of the \textit{Kepler Legacy} targets from \cite{lund17} and \cite{silva17}. Inferred parameters values are taken considering all the results from the methods described in \cite{silva17} (except inferred ages very far compare to the other methods).}

\begin{tabular}{l|cccc|cccc}
\noalign{\smallskip}\hline\hline\noalign{\smallskip}
&\multicolumn{4}{c|}{Observational constraints}&\multicolumn{4}{c}{Inferred parameters}\\
\noalign{\smallskip}\hline\noalign{\smallskip}
KIC & $T_\mathrm{eff}$~(K) & [Fe/H]~(dex) & $\nu_\mathrm{max}$ ($\mu$Hz)& $v\sin i$ (km~s$^{-1}$)&  Mass (M$_\odot$) & Radius (R$_\odot$) & Age (Gyr) & $X_C$  \\
\noalign{\smallskip}\hline\noalign{\smallskip}
2837475  & $6614 $  & $0.01 $ & $1558$ & $23.30$ & $[1.38;1.46]$  & $[1.61;1.65]$  & $[1.42;1.76]$  &  $[0.35;0.45]$\\
9206432  & $6538 $  & $0.16 $ & $1866$ & $6.80$  & $[1.38;1.60]$  & $[1.50;1.54]$  & $[1.20;1.96]$  &  $[0.37;0.50]$\\
11081729 & $6548 $  & $0.11 $ & $1969$ & $24.10$ & $[1.26;1.50]$  & $[1.42;1.49]$  & $[1.84;2.17]$  &  $[0.35;0.53]$\\
11253226 & $6642 $  & $-0.08$ & $1591$ & $14.40$ & $[1.32;1.46]$  & $[1.58;1.65]$  & $[1.60;1.88]$  &  $[0.34;0.39]$\\
\noalign{\smallskip}\hline\noalign{\smallskip}
\end{tabular}
\end{table*}

\textit{Kepler Legacy} stars KIC 2837475, KIC 9206432, KIC 11081729 and KIC 11253226 are stars potentially affected by inaccuracies in the stellar parameter inferences due to an improper modelling of transport processes in their radiative interiors. The stellar parameters for the four stars are listed in Table \ref{table:8}. All four stars have inferred masses between 1.32 and 1.46~M$_\odot$ with metallicities around the solar one. Their surface rotation velocities are smaller than 25 km/s and the range of $\nu_\mathrm{max}$ is in the region where abundances variations are predicted to be the most important (see Fig. \ref{fig:1}, \ref{fig:4}). This implies that the currently determined stellar parameters are possibly overestimated by [0.05,0.08]~M$_\odot$ ($\approx 3-5\%$) for the mass, 0.4~Gyr ($\approx 25\%$) for the age, and 0.04~R$_\odot$ ($\approx 2.5\%$) for the radius. Age and radius errors due to neglecting atomic diffusion (with radiative accelerations) and rotation are of the order of the dispersion obtained by \cite{silva17}, but the error on the age is much larger (except for KIC 9206432, which is around $20\%$). Inferring the parameters of these stars with a grid of models including atomic diffusion (with radiative accelerations) and rotation would then lead to significantly different ages. This of course strongly depends on the metallicity because the transport of chemical elements by atomic diffusion and rotation depends on the size of the surface convective zone. At a given mass, the size of the surface convective zone is not the same depending on the metallicity. The effect is then most likely larger when the metallicity is lower, at a given mass (i.e. larger for KIC 11253226 than KIC 9206432). A specific modelling of these four stars taking into account atomic diffusion and rotation will be performed as well as the individual modelling of stars in the three predicted regimes, in a forthcoming paper.

Another possibility to infer the parameters of these stars would be to use local optimization methods such as those based on the Levenberg-Marquardt minimization  \citep{2005A&A...441..615M,lebreton14}. Such methods have the advantage that they allow to easily explore larger ranges of possible input physics and parameters on-the-fly and they do not require to calculate a grid of models for each of them. However, they require some care, due to risks related to possible local minima and they cost more in computational time. They will be investigated in a forthcoming paper. 

\subsection{Impact of the mixing-length parameter}\label{alpha}

For all the models considered in this work we choose to keep the same $\alpha_\mathrm{CGM}$. This choice was made in order to be able to compare the models with exactly the same physics except for the processes (atomic diffusion, rotation) we wanted to test and to study their intrinsic impact.

In order to estimate the impact of keeping the same value for the mixing-length parameter, we computed a model of 1.4~M$_\odot$ (same as A8) with a solar calibrated $\alpha_\mathrm{CGM}=0.633$ (instead of $\alpha_\mathrm{CGM}=0.68$) and solar calibrated initial chemical composition. Then we inferred the parameters of this model using Grid1. We found a mass 0.06~M$_\odot$ smaller, a radius 0.02~R$_\odot$ smaller and an age 0.2~Gyr larger. It means that by not using the solar calibrated $\alpha_\mathrm{CGM}$ value  and initial chemical composition, we probably under-estimate masses and radii, and over estimate ages. In almost all the cases presented in Section \ref{resultAIMS}, we found that neglecting atomic diffusion (including radiative accelerations) and rotation leads to an over-estimation of masses, radius and ages when using Grid1. The net effect when using solar calibrated values in accordance with the physical assumptions is expected to be an even larger overestimate of the mass and a smaller one for the age when neglecting diffusion and rotation. We note however that even in this case, our predicted impact on the age is larger than the one coming from the different mixing-length parameter values.

We must also stress that there is no reason a priori to use the solar calibrated values for stars more massive than the Sun. The mixing-length parameters was indeed found to vary with evolution and masses \citep[for instance][]{Freytag99, trampedach14,sonoi19}. \cite{sonoi19} showed that the $\alpha_\mathrm{CGM}$ value can vary from 0.583 to 0.801 for stars with $T_\mathrm{eff}$=[4018;6725]~K and $\log g$=[1.50;4.50]~dex while the calibrated solar value was 0.69. For example, their computations of the mixing-length parameter for a model close to our A8 model gives $\alpha_\mathrm{CGM}$=0.636 (see their Table A1).

To minimize this source of uncertainties, we plan to build a grid of models including atomic diffusion (including radiative accelerations), rotation and taking into account various values of $\alpha_\mathrm{CGM}$ and the overshoot parameter as well.

\section{Conclusion}

We studied the impact of taking both atomic diffusion and rotationally induced mixing into account on surface abundances and inferences of stellar parameters such as mass, age and radius for main sequence stars with masses in the range [1.3-1.44] $M_\odot$. Atomic diffusion here includes radiative accelerations, and rotation is assumed either differential in radius or uniform. We showed that a strong interaction between atomic diffusion and rotation exists throughout the size of the convective envelope. Indeed the accumulation of elements at the surface leads to an increase of the surface convective zone due to a local increase of the opacity. As the surface convective zone is larger, it is able to maintain a stronger magnetic field by dynamo. This generates a more efficient extraction of angular momentum by magnetized wind, and a stronger mixing induced by rotation which in turn tends to decrease the effect of atomic diffusion.

We showed that the efficiency of the transport of chemical elements is strongly related to the efficiency of the transport of angular momentum. Atomic diffusion and rotation are then in competition. We considered a grid of standard stellar models. Here, standard means convection but no transport in radiative regions. Comparison of these models with models including rotationally induced mixing and/or atomic diffusion led us to distinguish three regimes depending on the mass of the star.

For stars with masses below and close to 1.3~M$_\odot$, when only atomic diffusion is taken into account, inferred ages depart significantly from the actual ones. This is mainly due to the surface [Fe/H], which differs from the initial one (up to 0.15~dex). This large difference in age almost totally vanishes when shellular rotation is also taken into account. Indeed the mixing induced by the rotation prevents large variations of surface abundances (helium and metals). The surface abundances are then consistent with observations. The efficiency of the rotationally induced mixing is due to an efficient extraction of angular momentum by magnetized wind, possibly because the (atomic diffusion-induced larger) size of the surface convective zone is sufficient to maintain a magnetic field. In this mass range, when rotation is taken into account, radiative accelerations have a very small effect on surface abundances. The deviation in mass/radius comes from the effect of rotation on the structure of the star. We must stress that the behaviour of surface abundances in this range of mass is expected to be strongly affected by additional angular momentum transport processes (i.e for instance internal gravity waves).

For stars with masses close to 1.4~M$_\odot$, when only atomic diffusion is taken into account, the effect on the inferred age is the same as the 1.3~M$_\odot$ case. As the surface convection is thinner, the extraction of angular momentum by magnetized wind is less efficient. The mixing efficiency of rotation is weaker. Hence when both atomic diffusion and rotation are considered, surface abundance variations are maintained unlike for the lower mass case. The [Fe/H] variation reaches 0.1~dex, and helium is depleted down to $\approx 0.20$ which is still consistent with observations. As [Fe/H] is lower than the initial one during the evolution on the main-sequence, the age inference deviates from the real value by as much as 0.4~Gyr when using a standard physics grid. This is roughly 20-25\% of the age when considering \textit{Kepler} stars (see Section \ref{kepler}). In this mass range, radiative accelerations are crucial and have a significant effect on surface abundances (0.1~dex) and on stellar parameter inference.

For stars with masses larger than 1.4~M$_\odot$, surface abundance variations are quite large with or without rotation. Iron is accumulated at the surface due to radiative accelerations. Without rotation, it is not possible to find a solution using a standard grid of stellar models. When the effect of rotation is included, inference of stellar parameters using a standard physics grid leads to important age deviations compared with the real value due to these large [Fe/H] differences. Radiative accelerations are crucial for parameter inference as rotation is no longer able to prevent the strong surface abundance depletion/accumulation. Surface abundances are no longer representative of "normal" F-type stars, but representative of Fm chemically peculiar stars, in the range of initial surface rotation speeds ($<100$~km/s) we considered. This is in disagreement with the surface abundance determinations of \textit{Kepler} stars. One must then conclude that there is a missing transport process of chemical elements in addition to those of atomic diffusion and rotationally induced mixing as modelled nowadays.

All the comparisons of this study were done at constant $X_C$. It is known that some effects may be compensated when comparing at constant $\nu_\mathrm{max}$ for instance. The results may slightly differ when performing a full optimization with a stellar model grid including atomic diffusion and rotation. The next step of the study is to perform such optimizations. We also plan to compare the results given by another optimization method (based on the Levenberg-Marquardt algorithm) in order to test the robustness of our conclusions. It will also allow us to test the impact of using a local method instead of a global one, as AIMS for instance.

Note that the above conclusions concerned stellar models with solar metallicity. For lower metallicities, the mass thresholds are smaller. 

We stress also that in most of the present study shellular rotation was assumed. However observations show that the actual transport of angular momentum implemented in evolutionary codes is underestimated. Nowadays a consensus seems to hold where stars are likely rotating uniformly rather than differentially radially in their radiative regions. We therefore considered also a uniform rotation which led to a small transport of chemical elements, hence to a stronger impact of atomic diffusion as early as 1.3~M$_\odot$. In that case, unlike shellular rotation, metals and helium are not compatible with observation. Both missing transport processes of angular momentum and of chemical elements are potentially related. Despite the current uncertainties about the modelling of angular momentum transport, which also have consequences on the transport of chemical elements, the three regimes should exist even if the modelling of these processes is improved. The main impact will be the range of mass for which each regime occurs.

One signature of transport of chemical elements in stars is the surface abundance of lithium. It traces the mixing occurring from the surface down to the region where lithium is destroyed by nuclear reactions ($T=2.5\times10^6$~K). The lithium abundances of stellar models including rotationally induced mixing  with shellular rotation are consistent with the hotter part of the lithium dip, as already emphasized long ago \citep[e.g.][]{talon08}. A more detailed study about the impact of the coupling between atomic diffusion (with radiative acceleration) and rotation on lithium is postponed to a forthcoming paper.

\begin{acknowledgements}
We thank Aldo Serenelli for fruitful discussions. We thank the referee Patrick Eggenberger for valuable comments which helped to improve the paper. This work was supported by FCT/MCTES through national funds (PIDDAC) by this grant UID/FIS/04434/2019 and by FEDER - Fundo Europeu de Desenvolvimento Regional through COMPETE2020 - Programa Operacional Competitividade e Internacionalizaç\~ao by these grants: UID/FIS/04434/2019; PTDC/FIS-AST/30389/2017 \& POCI-01-0145-FEDER-030389. MD is supported in the form of a work contract funded by national funds through Fundação para a Ciência e Tecnologia (FCT). This work was also supported by the CNES. We acknowledge financial support from the "Programme National de Physique Stellaire" (PNPS) of the CNRS/INSU co-funded by the CEA and the CNES, France.
 
\end{acknowledgements}  

%%%%%%%%%%%%%%%%%%%%%%%%%%%%%%%%%%%%%%%%%%%%%%%%%%%%%%%%%%%%%%%%%%%  
\bibliographystyle{aa} % style aa.bst
\bibliography{rota_vs_diff.bib} % your references Yourfile.bib  
%%%%%%%%%%%%%%%%%%%%%%%%%%%%%%%%%%%%%%%%%%%%%%%%%%%%%%%%%%%%%%%%%%%
%_____________________________________________________________________

\end{document}